%% file: ICdelay_1col.tex
\documentclass[11pt, draftclsnofoot, onecolumn]{IEEEtran}

\usepackage{graphicx}
\usepackage{subfigure}
\usepackage{caption2}
\usepackage{amssymb}
\usepackage{amsthm}
\usepackage{amsmath}
\usepackage{mathtools}
\usepackage{cite}
\begin{document}
%
\title{Achievable Sum DoF of the K-User MIMO Interference Channel with Delayed CSIT}

\author{Chenxi Hao and Bruno Clerckx 
\thanks{Chenxi Hao and Bruno Clerckx are with the Communication and Signal Processing group of Department of Electrical and Electronic Engineering, Imperial College London. Bruno Clerckx is also with the School of Electrical Engineering, Korea University. This paper has been partially presented in \cite{Hao15MISOIC}.}
}

\maketitle

\begin{abstract}
This paper considers a $K$-user multiple-input-multiple-output (MIMO) interference channel (IC) where 1) the channel state information obtained by the transmitters (CSIT) is completely outdated, and 2) the number of transmit antennas at each transmitter, i.e., $M$, is greater than the number of receive antennas at each user, i.e., $N$. The usefulness of the delayed CSIT was firstly identified in a $K$-phase Retrospective Interference Alignment (RIA) scheme proposed by Maddah-Ali et al for the Multiple-Input-Single-Output Broadcast Channel, but the extension to the MIMO IC is a non-trivial step as each transmitter only has the message intended for the corresponding user. Recently, Abdoli et al focused on a Single-Input-Single-Output IC and solved such bottleneck by inventing a $K$-phase RIA with distributed overheard interference retransmission. In this paper, we propose two $K$-phase RIA schemes suitable for the MIMO IC by generalizing and integrating some key features of both Abdoli's and Maddah-Ali's works. The two schemes jointly yield the best known sum Degrees-of-Freedom (DoF) performance so far. For the case $\frac{M}{N}{\geq}K$, the achieved sum DoF is asymptotically given by $\frac{64}{15}N$ when $K{\to}\infty$.
\end{abstract}
\newtheorem{mytheorem}{Theorem}
\newtheorem{mycoro}{Corollary}
\newtheorem{myremark}{Remark}
\newtheorem{mydef}{Definition}
\newtheorem{myassump}{Assumption}
\section{Introduction}\label{sec:Intro}
\input{Intro_ICdelay_1col}

\section{System Model}\label{sec:SM}
\input{SM_ICdelay_1col}

\section{Main Results}\label{sec:MR}
\input{MR_ICdelay_1col}

\section{Achievable scheme in the $(K{,}1{,}K)$ IC}\label{sec:K1K}
\input{ach_K1K}

\section{Achievable schemes in the $(M{,}N{,}K)$ IC}\label{sec:MNK}
\input{ach_MNK}

%
%

\section{Conclusion}\label{sec:conclusion}
This paper considers a $(M{,}N{,}K)$ IC with perfect delayed CSIT, where each transmitter has $M$ antennas and each user has $N$ antennas and $M{\geq}N$. Building upon the $K$-phase RIA with distributed overheard interferences retransmission, we propose two schemes suitable for the $(M{,}N{,}K)$ IC by generalizing and integrating the key ingredients of redundancy transmission, partial interference nulling proposed by Abdoli et al and the MAT-like transmission proposed by Maddah-Ali et al. Moreover, we consider that both schemes are performed via a transmitter-user pairs scheduling in phase $1$. By finding the optimal number of co-scheduled transmitters, the two schemes jointly achieve a greater sum DoF performance than all the previously known results for the general $(M{,}N{,}K)$ IC with $M{\geq}N$.

%

\section*{Appendix}
\input{AppICdelay_1col}

\bibliographystyle{IEEEtran}

\bibliography{ICdelay}

\end{document}

%% file: Intro_ICdelay_1col.tex
The capacity region of the multiple-input-multiple-output (MIMO) interference channel (IC) has remained an open problem for decades. However, during the past decade, there have been extensive researches on another important system metric, i.e., Degrees-of-Freedom (DoF), as it sheds light on the behavior of the capacity at high Signal-to-Noise-Ratio (SNR). The sum DoF or DoF region of the IC was studied in \cite{JafarIA08,GouKMN,Wang14SubspaceChain,JarfarMIMOIA2pair} for the case with perfect channel state information at the transmitters (CSIT), and was studied in \cite{VV11a,HJSV09,zhunoCSIT} for the case with no CSIT. In practical systems, the CSIT can be outdated due to large propagation delay and high user mobility. When the latency is comparable to the channel coherence time, the sum DoF achieved by conventional multi-user schemes, such as interference alignment, degrades dramatically and is no better than the case with no CSIT \cite{Ges12}. Therefore, inventing novel transmission strategies that make use of the delayed CSIT has attracted many researchers. For convenience, in the rest of the paper, when we mention a $(M{,}N{,}K)$ IC (resp. BC), it means that there are $K$ transmitter-user pairs (resp. $K$ users associated with the single transmitter), $M$ antennas at each transmitter (resp. the single transmitter), and $N$ antennas at each user.

The usefulness of the outdated CSIT was firstly found by Maddah-Ali and Tse in \cite{Tse10}, focusing on a $(K{,}1{,}K)$ broadcast channel (BC). The key idea is as follows. With outdated CSIT, the transmitter can reconstruct the previously overheard interference at various users to create future transmissions. These future transmissions provide additional useful signals for some users while aligning previously overhead interferences for some other users. Using this philosophy, the authors invented a $K$-phase Retrospective Interference Alignment (RIA) with centralized overheard retransmission (known as MAT scheme). This scheme achieves the optimal sum DoF $\frac{K}{1{+}\frac{1}{2}{+}{\cdots}{+}\frac{1}{K}}$ in the $(K{,}1{,}K)$ BC, outperforming the case with no CSIT. An alternative two-user MAT scheme was also proposed in \cite{Tse10,Ges12} for the two-user case, which differs by the way the overheard interferences are generated and retransmitted. Moreover, the diversity-multiplexing trade-off achieved by the MAT and alternative MAT schemes were reported in \cite{Bruno15}, while the integration of the alternative MAT scheme and statistical beamforming was studied in \cite{Mingbo15}.

However, the $K$-phase RIA with centralized overheard retransmission is not generally applicable to the $(K{,}1{,}K)$ IC except for the two-user case \cite{TorrellasK,VVICdelay,xinping_miso_ic}. This is because without data-sharing each transmitter cannot reconstruct the whole interference at a user when the interference originates from more than one interferer. Due to this fact, some works including \cite[Theorem 5]{Ghasami11} and \cite{TorrellasK} for the $(K{,}1{,}K)$ IC, and \cite{MalekiRIA,Maggi12} for the $(1{,}1{,}K)$  Single-Input-Single-Output (SISO) IC, designed $2$-phase schemes. In \cite{Ghasami11,TorrellasK}, the overheard interference resulted at the end of phase $1$ is retransmitted one-by-one, i.e., following a time-sharing fashion. Essentially, in the philosophy of the $K$-phase RIA scheme, the delivery of the overheard interferences generated in phase $1$ is accomplished through phase $2$ to $K$ by making use of the perfect delayed CSIT. Therefore, the $K$-phase RIA scheme is likely to achieve a greater sum DoF performance than those $2$-phase schemes. This fact make it appealing to look for a $K$-phase RIA scheme for the $(K{,}1{,}K)$ IC. In \cite{Abdoli13}, Abdoli et al invented a $K$-phase RIA with \emph{distributed} overheard interference retransmission focusing on a $(1{,}1{,}K)$ IC. Compared to the MAT scheme designed for BC, Abdoli's scheme features a \emph{distributed higher order symbol generation} and a transmitter scheduling in phase $2$ through to phase $K$ (to be introduced later on). For the $(1{,}1{,}3)$ IC, the achievable sum DoF is $\frac{36}{31}$, outperforming $\frac{9}{8}$ achieved by the two-phase scheme in \cite{MalekiRIA,Maggi12}.


Moreover, in the context of $(M{,}N{,}K)$ IC, recent works\cite{TorrellasMNK} attempted to generalize the schemes proposed in \cite{TorrellasK,MalekiRIA,Abdoli13}. However, their schemes have the following two drawbacks: 1) only $2$-phase and $3$-phase schemes are applied in the $K$-user case, so that the benefit of delayed CSIT in performing RIA is not fully exploited; 2) when generalizing the scheme designed for the $(1{,}1{,}3)$ IC in \cite{Abdoli13} to the $(M{,}N{,}3)$ IC with $M{\geq}N$, the sum DoF is $\frac{36}{31}N$, which is simply a scaled version of the $(1{,}1{,}3)$ IC, revealing a wasteful use of the extra transmit antennas. Therefore, the usefulness of the $K$-phase RIA framework proposed in \cite{Abdoli13} has only been properly identified in the $(1{,}1{,}K)$ IC thus far, leaving aside the questions 1) whether it is applicable to the $(M{,}N{,}K)$ IC, and 2) how the transmission strategy in each phase changes with $M$ and $N$. These are the main focuses of this paper. Our main contributions are highlighted as follows.

Firstly, we propose a transmission scheme for the $(K{,}1{,}K)$ IC by integrating the $K$-phase RIA framework proposed in \cite{Abdoli13} and the key features of the MAT-like transmission \cite{Tse10} in each phase.

Secondly, building on the $K$-phase RIA framework proposed in \cite{Abdoli13}, we propose two achievable schemes suitable for the $(M{,}N{,}K)$ IC with $1{\leq}\frac{M}{N}{\leq}K$, which are generalizations of the MAT-like transmission designed for the $(K{,}1{,}K)$ IC in the first contribution and the redundancy transmission and partial interference nulling (RT-PIN) approach proposed in \cite{Abdoli13}. The details of the novelties of the proposed schemes are presented in Section \ref{sec:MR}.

\begin{figure*}[t]
\renewcommand{\captionfont}{\small}
\captionstyle{center}
\centering
\subfigure[$(1{,}1{,}K)$ IC: Proposed scheme vs. Abdoli's scheme \cite{Abdoli13}]{
                \centering
                \includegraphics[width=0.25\textwidth,height=3cm]{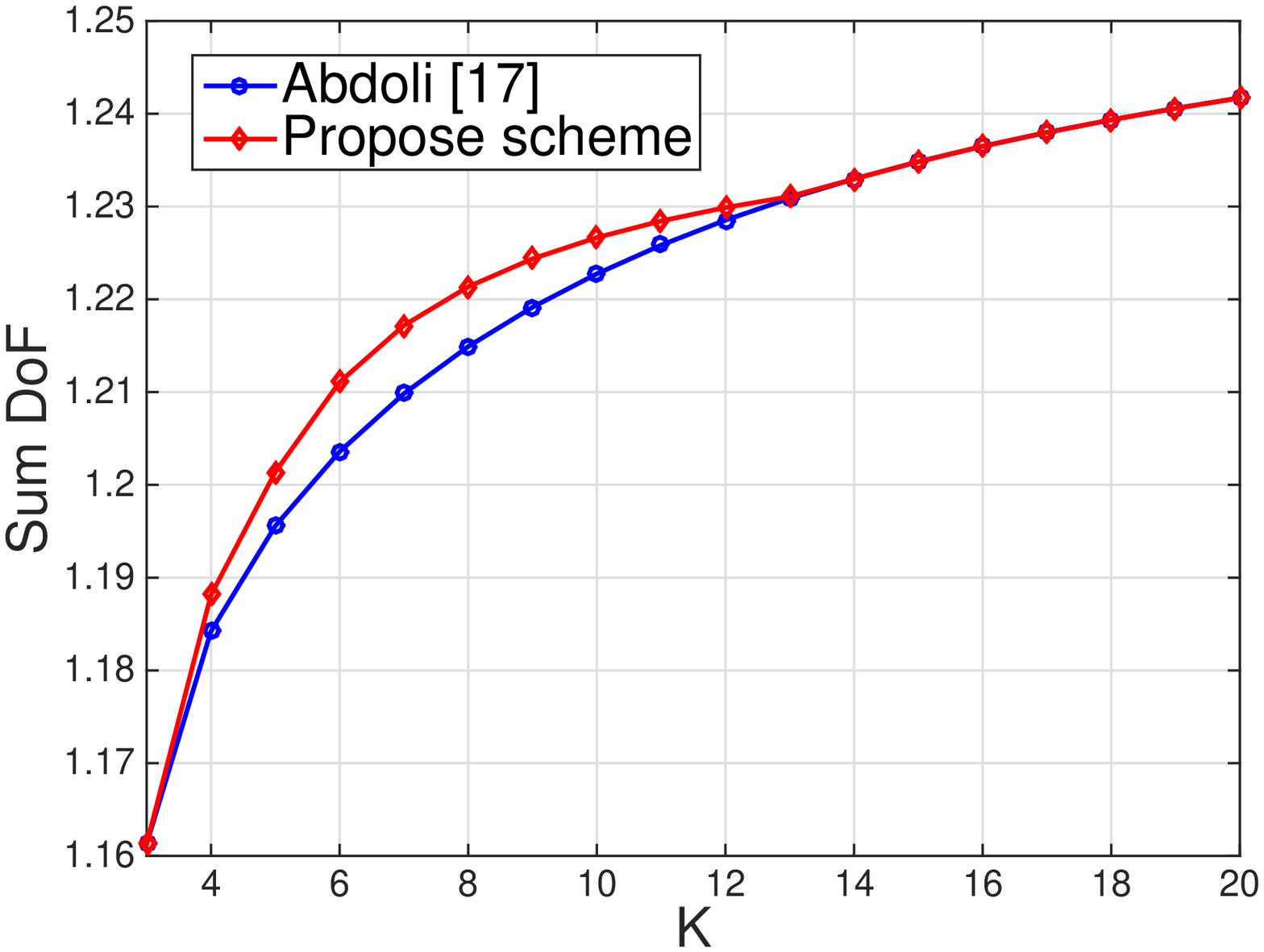}
                \label{fig:SumDoF11K}
        }
\subfigure[$(M{,}N{,}3)$ IC]{
                \centering
                \includegraphics[width=0.25\textwidth,height=3cm]{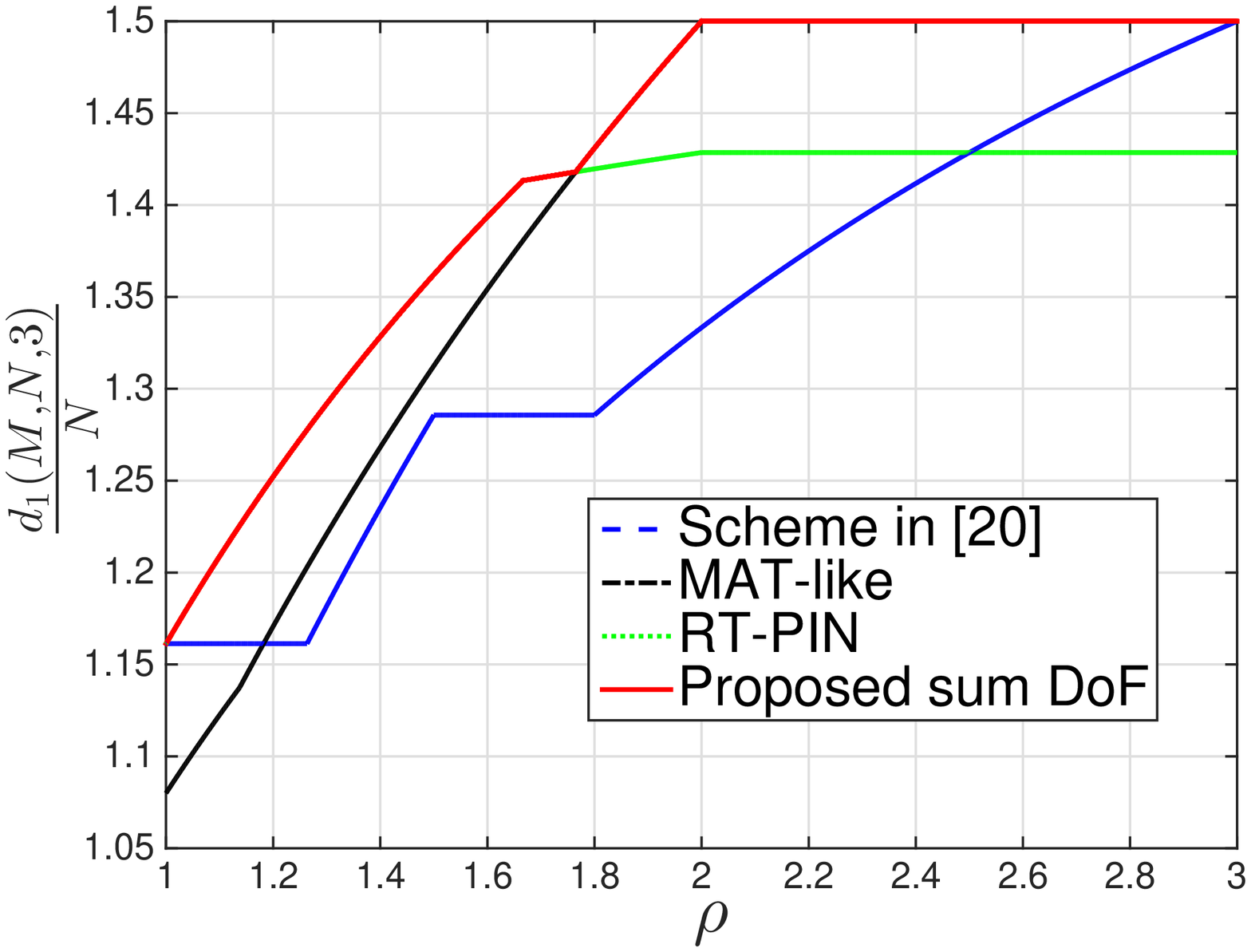}
                \label{fig:K3}
        }
\subfigure[$(M{,}N{,}6)$ IC]{
                \centering
                \includegraphics[width=0.25\textwidth,height=3cm]{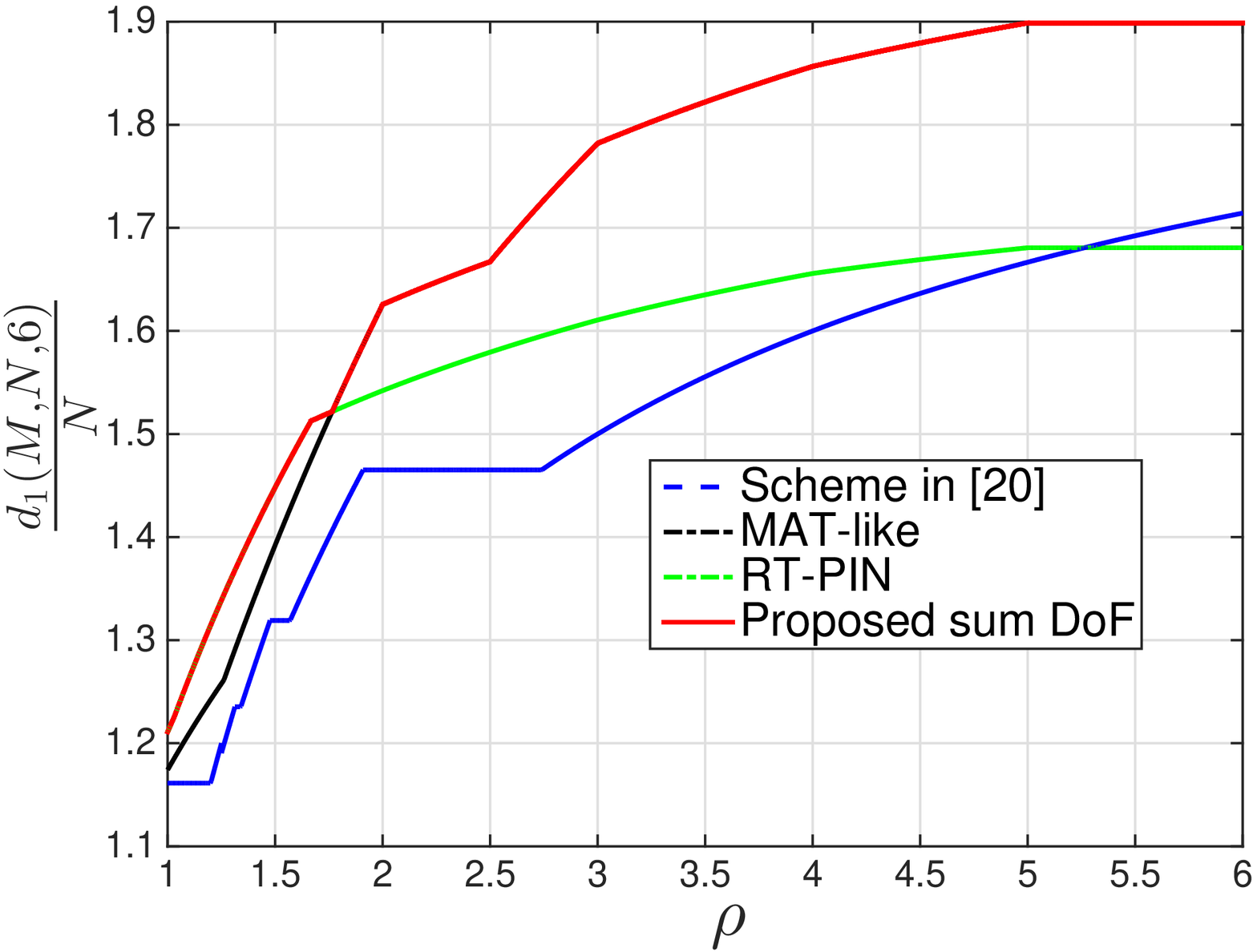}
                \label{fig:K6}
        }
\subfigure[$(K{,}1{,}K)$ IC]{
                \centering
                \includegraphics[width=0.25\textwidth,height=3cm]{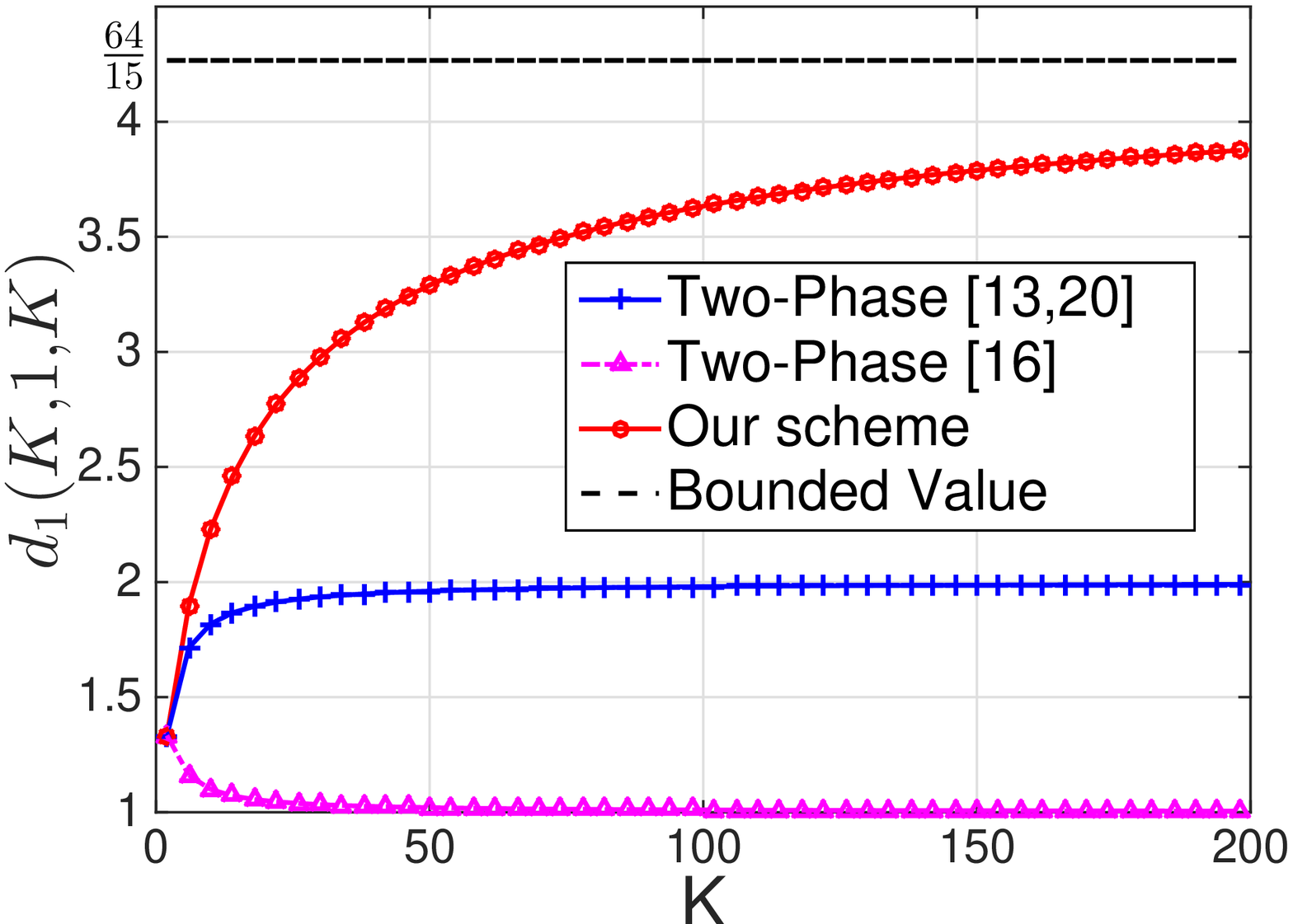}
                \label{fig:MISO}
        }
\caption{Normalized achievable sum DoF, i.e., $\frac{\text{\rm sum DoF}}{N}$, where $\rho{\triangleq}\frac{M}{N}$.}\label{fig:ds}
\end{figure*}

Thirdly, in both schemes, we consider that there are $n$ (out of $K$) transmitters active in each slot of phase $1$ delivering private symbols to their corresponding users, i.e., so-called $n$-transmitter/$n$-user scheduling. We obtain the achievable sum DoF by 1) taking the maximum of the sum DoF achieved by the two schemes mentioned in the second contribution, and 2) finding the optimal number of co-scheduled transmitters. As shown in Figure \ref{fig:SumDoF11K}, the scheduling process allows us to improve the sum DoF of the $(1{,}1{,}K)$ IC achieved by the scheme proposed in \cite{Abdoli13}. Moreover, as illustrated in Figure \ref{fig:K3} for the $(M{,}N{,}3)$ IC, Figure \ref{fig:K6} for the $(M{,}N{,}6)$ IC, and Figure \ref{fig:MISO} for the $(K{,}1{,}K)$ IC, our scheme (red solid curve) significantly outperforms all the previously known results. Besides, in the $(K{,}1{,}K)$ IC, the achievable sum DoF tends to $\frac{64}{15}$ when $K{\to}\infty$.

Organization: Section \ref{sec:SM} elaborates on the system model. Section \ref{sec:MR} summarizes the key ingredients of the proposed schemes and lists the theorems on achievable sum DoF. The achievable scheme designed for $(K{,}1{,}K)$ IC and the achievable schemes designed for the general $(M{,}N{,}K)$ IC with $1{\leq}\frac{M}{N}{\leq}K$ are presented in Section \ref{sec:K1K} and Section \ref{sec:MNK}, respectively. Section \ref{sec:conclusion} concludes the paper.

Notations: Bold lower letters stand for vectors whereas a symbol not in bold font represents a scalar. $\left({\cdot}\right)^T$ and $\left({\cdot}\right)^H$ denote the transpose and the Hermitian of a matrix or vector, respectively. ${\parallel}{\cdot}{\parallel}$ is the norm of a vector. The block diagonalization of matrices $\mathbf{A}$ of size $m{\times}n$ and $\mathbf{B}$ of size $p{\times}r$ is referred as $\text{\rm Bdiag}\{\mathbf{A}{,}\mathbf{B}\}{\triangleq}\left[\begin{array}{cc}\mathbf{A} & \mathbf{0}_{m{\times}r}\\ \mathbf{0}_{p{\times}n} & \mathbf{B}\end{array}\right]$. $\mathbb{E}\left[{\cdot}\right]$ refers to the statistical expectation. $|\mathcal{S}|$ is the cardinality of the set $\mathcal{S}$. ${\lfloor}{a}{\rfloor}$ and ${\lceil}a{\rceil}$ stand for the greatest integer that is smaller than $a$ and the smallest integer that is greater than $a$, respectively. The linear space spanned by the rows of matrix $\mathbf{A}$ is denoted by $\text{\rm rowspan}\{\mathbf{A}\}$. The stack of two matrices with the same number columns writes as ${\rm stack}\{\mathbf{A}{,}\mathbf{B}\}{\triangleq}\left[\begin{array}{c}\mathbf{A}\\ \mathbf{B}\end{array}\right]$.

%% file: SM_ICdelay_1col.tex
In the $(M{,}N{,}K)$ IC with $M{\geq}N$, the received signal $\mathbf{y}_k(t){\in}\mathbb{C}^{N{\times}1}$, in a certain time slot $t$ writes as
\begin{IEEEeqnarray}{rcl}
\mathbf{y}_k(t)&{=}&\sum_{j{\in}\mathcal{S}_K}\mathbf{H}_{kj}(t)\mathbf{s}_j(t){+}\mathbf{z}_{k}(t),\forall k{\in}\mathcal{S}_K{,}
\end{IEEEeqnarray}
where $\mathcal{S}_K{\triangleq}\{1{,}\cdots{,}K\}$ and $\mathbf{z}_{k}(t)$ represents the Additive White Gaussian Noise with zero mean and unit variance. We consider that the transmitted signal $\mathbf{s}_j(t){\in}\mathbb{C}^{M{\times}1}$ is subject to the power constraint $P$, i.e., $\mathbb{E}[{\parallel}\mathbf{s}_j(t){\parallel}^2]{\leq}P$. The matrix $\mathbf{H}_{kj}(t)$, of size ${N{\times}M}$, refers to the channel between transmitter $j$ (Tx$j$) and user $k$ (Rx$k$). $\mathbf{H}_{kj}$ has circularly symmetric complex Gaussian entries with zero mean and unit variance (Rayleigh fading). The fading process is i.i.d across time slots (fast fading) and links.

In wireless systems like Long Term Evolution, in Time Division Duplexing mode, the CSI is measured on the uplink and used in the downlink assuming channel reciprocity. In Frequency Division Duplexing mode, each user estimates their CSI using pilot and the estimated CSI is quantized and reported to its serving transmitter via a rate-limited feedback link. Furthermore, in both modes, in order to perform multi-cell coordination and/or joint transmission, the CSI has to be shared among the transmitters via a backhaul link. Due to the latency incurred in the feedback link and backhaul link, and because of the user mobility, the CSI acquired by the transmitters is relatively out-dated compared to the current CSIT. In this paper, to investigate the usefulness the delayed CSIT in the $K$-user interference channel, we consider a general setup where each transmitter acquires the global CSI with one-slot delay, while each user has perfect knowledge of the global current CSI to perform the decoding. Specifically, at the beginning of a certain time slot $t$, each Tx perfectly knows $\mathbf{H}_{jk}(1){,}{\cdots}{,}\mathbf{H}_{jk}(t{-}1){,}{\forall}{j{,}k}{=}1{,}{\cdots}{,}K$, while each Rx perfectly knows $\mathbf{H}_{jk}(1){,}{\cdots}{,}\mathbf{H}_{jk}(t){,}{\forall}{j{,}k}{=}1{,}{\cdots}{,}K$. However, at the end of slot $t$, each Tx obtains $\mathbf{H}_{jk}(t){,}{\forall}{j{,}k}{=}1{,}{\cdots}{,}K$ as well.

Let us consider that Tx$k$ has a private message, i.e., $c[k]$, for Rx$k$. Then, the rate $R[k]$ is achievable if Rx$k$ decodes $c[k]$ with arbitrary small error probability as the codeword length approaches infinity. Consequently, the system metric, namely sum DoF, is given by
\begin{IEEEeqnarray}{rcl}
d_1(M{,}N{,}K)&{\triangleq}&\lim_{P\to\infty}\frac{\sum_{k{=}1}^KR[k]}{{\log}P}.\label{eq:sumdof}
\end{IEEEeqnarray}

In this paper, we use $\mathcal{S}_K$ to denote the set of the $K$ users. Besides, similar to that defined in \cite{Tse10,Abdoli13}, we introduce following definitions, which are frequently used in the subsequent derivations and analysis.
\begin{mydef}\label{def:orderm}
Considering that for any subset of $m$ ($1{\leq}m{\leq}K$) transmitter-user pairs, i.e., $\mathcal{S}_m{\subseteq}\mathcal{S}_K$ and $|\mathcal{S}_m|{=}m$, if any transmitter $k$, $k{\in}\mathcal{S}_m$, has a message, $w[k{|}\mathcal{S}_m]$, intended for all the users in $\mathcal{S}_m$, then we term such a message as an order-$m$ message.
\end{mydef}
\begin{mydef}\label{def:order1m}
Considering that for any subset of $m{+}1$ ($2{\leq}m{\leq}K{-}1$) transmitter-user pairs, i.e., $\mathcal{S}_{m{+}1}{\subseteq}\mathcal{S}_K$ and $|\mathcal{S}_{m{+}1}|{=}m{+}1$, if any transmitter $k$, $k{\in}\mathcal{S}_{m{+}1}$, has a message, $w[k{|}k{;}\mathcal{S}_{m{+}1}{\setminus}k]$, intended for user $k$, but already known by other $m$ users in $\mathcal{S}_{m{+}1}{\setminus}k$, then we term such a message as an order-$(1{,}m)$ message.
\end{mydef}


The rate $R[k{|}\mathcal{S}_m]$, is said to be achievable if all the users in $\mathcal{S}_m$ decode the order-$m$ message $w[k{|}\mathcal{S}_m]$ with arbitrary small error probability as the codeword length approaches infinity. Similarly, the rate $R[k{|}k{;}\mathcal{S}_{m{+}1}{\setminus}k]$ is said to be achievable if user $k$ decodes the order-$(1{,}m)$ message $w[k{|}k{;}\mathcal{S}_{m{+}1}{\setminus}k]$ with arbitrary small error probability as the codeword length approaches infinity. Then, the sum DoF of delivering all the order-$m$ messages (${\forall}2{\leq}m{\leq}K$) and all the order-$(1{,}m)$ messages (${\forall}2{\leq}m{\leq}K{-}1$), are given by
\begin{align}
d_m(M{,}N{,}K){\triangleq}&\lim_{P\to\infty}
\frac{\sum_{{\forall}k{\in}\mathcal{S}_m{,}{\forall}\mathcal{S}_m{\subseteq}\mathcal{S}_K}R[k{|}\mathcal{S}_m]}{{\log}_2P},\label{eq:sumdofm}\\
d_{1{,}m}(M{,}N{,}K){\triangleq}&\lim_{P\to\infty}
\frac{\sum_{{\forall}k{\in}\mathcal{S}_{m{+}1}{,}{\forall}\mathcal{S}_{m{+}1}{\subseteq}\mathcal{S}_K}R[k{|}k{;}\mathcal{S}_{m{+}1}{\setminus}k]} {{\log}_2P}{,}
\end{align}
respectively. When $m{=}1$, \eqref{eq:sumdofm} becomes the sum DoF of the private messages as defined in \eqref{eq:sumdof}.

Furthermore, we reuse the notation in \cite{Abdoli13}, i.e., $u[k|\mathcal{S}_m;\mathcal{S}_{m^\prime}]$, to represent a symbol which is 1) transmitted by Tx$k$, 2) desired by a subset $\mathcal{S}_m$ of users, where $|\mathcal{S}_m|{=}m$, and 3) already known by a subset $\mathcal{S}_{m^\prime}$ of users, where $|\mathcal{S}_{m^\prime}|{=}m^\prime$ and $\mathcal{S}_m{\cap}\mathcal{S}_{m^\prime}{=}\emptyset$. With such a notation, we introduce two classes of symbols:
\begin{itemize}
\item Order-$m$ symbols (formed by the order-$m$ message $c[k{|}\mathcal{S}_m]$), denoted by $u[k|\mathcal{S}_m]$, which is desired by a subset $\mathcal{S}_m$ of users, and known by no user, i.e., $\mathcal{S}_{m^\prime}{=}\emptyset$;
\item Order-$(1{,}m^\prime)$ symbols, denoted by $u[k|k;\mathcal{S}_{m^\prime}]$, which is intended for one user i.e., Rx$k$, but already known by other $m^\prime$ users, where $k{\notin}\mathcal{S}_{m^\prime}$.
\end{itemize}
For convenience, an order-$1$ symbol is denoted by $u_k$ (short for $u[k{|}k]$). It carries the private message of Rx$k$, and is only desired by Rx$k$, but is unknown to all users. As we will see later on, the symbols sent at the beginning of the communication are order-$1$ symbols. The symbols transmitted in phase $m$-I, $2{\leq}m{\leq}K$, are regarded as order-$m$ symbols as they are useful to a certain subset of $m$ users, while the symbols transmitted in phase $m{+}1$-II, $2{\leq}m{\leq}K{-}1$, are order-$(1{,}m)$ symbols as they are to be decoded by only one user and already known by another $m$ users. 

%% file: MR_ICdelay_1col.tex
In this paper, we design $K$-phase RIA schemes based on the framework proposed in \cite{Abdoli13}. Specifically, a distributed overheard interference retransmission is performed in each phase, and the transmitted signal in the next phase is built via a distributed higher order symbol generation. Besides, the novelties of the transmission block in each phase is highlighted as follows.
\subsubsection{Phase 1, MAT-like transmission}
We firstly focus on the $(K{,}1{,}K)$ MISO IC. In this case, as the number of transmit antennas is large enough, the overheard interference obtained at various unintended users are linearly independent of each other. Thus, the MAT scheme designed for the MISO BC can be reused. Secondly, we consider the case $1{\leq}\frac{M}{N}{<}K$. In this case, since the overheard interference at various users are linearly dependent, we modify the MAT scheme by exploiting part of the overheard interference as useful signals to create future transmissions.
\subsubsection{Phase 1, Two-Stage RT and PIN}
The redundancy transmission (RT) was firstly introduced for the $(1{,}1{,}K)$ IC. Specifically, each Tx delivers $t^\prime$ symbols in $t$ slots, where $t^{\prime}{<}t$. As the number of receive antennas is equal to the number of transmit antennas, each overheard interference spans a subspace of the received signal, thus allowing each user to null out the interference originated from one interferer (partial interference nulling, PIN). To enable PIN in the MIMO case $1{\leq}\frac{M}{N}{\leq}K$, the number of transmitted symbols has to be smaller than the number of receive antennas. To overcome this bottleneck, we design a two-stage redundancy transmission, where the first stage is used for interference sensing, while the second stage is designed using delayed CSIT to force the interference into the linear space created in the first stage. As we will see later on, this two-stage transmission allows the number of transmitted symbols scale with the number of transmit antennas.
\subsubsection{Phase 1, Tx-Rx pair scheduling}
For both MAT-like transmission and Two-stage RT and PIN, we perform a Tx-Rx pair scheduling in phase $1$ to improve the sum DoF performance.
\subsubsection{Phase 2 through to Phase $K$}
The scheme generalizes the RT-PIN approach proposed in \cite{Abdoli13}, where there are two active transmitters per slot and the pair of active transmitters are scheduled in a cyclic order. The novelty lies in that the number of order-$m$ symbols transmitted per slot is properly determined according to the number of antennas at each transmitter and each user. When $M{\geq}N(K{-}m{+}1)$, the scheme smoothly connects with the MAT-like transmission, where there is only one active user and the overheard interference at various users is directly exploited as useful signals without the need of PIN.

According to the definitions and assumptions made in Section \ref{sec:SM}, we state the main results on the achievable sum DoF as follows.
\begin{mytheorem}\label{theo:K1K}
For a $(K{,}1{,}K)$ IC ($K{\geq}2$) with perfect completely outdated CSIT, an achievable sum DoF by integrating MAT scheme with distributed overheard interference retransmission is given by
\begin{IEEEeqnarray}{rcl}
d_1(K{,}1{,}K){=}\max_{i{=}1{,}2}\frac{\mathcal{O}_i(K)^2}{1{+}\frac{\mathcal{O}_i(K)(\mathcal{O}_i(K){-}1)}{d_2(K{,}1{,}K)}}{,}\text{\rm almost surely,}\label{eq:d1_K1K}
\end{IEEEeqnarray}
where $\mathcal{O}_1(K){=}\lfloor2d_2(K{,}1{,}K)\rfloor$, $\mathcal{O}_2(K){=}\lceil2d_2(K{,}1{,}K)\rceil$ and
\begin{IEEEeqnarray}{rcl}
d_2(K{,}1{,}K){=}\left[1{-}\frac{1}{K{-}1}\sum_{l{=}2}^{K{-}1}\frac{K{-}l}{l^2{-}1}\right]^{-1}.\label{eq:d2_K1K}
\end{IEEEeqnarray}
refers to the sum DoF achieved by delivering order-$2$ symbols.
\end{mytheorem}

\begin{myremark}
\emph{Theorem \ref{theo:K1K} can be easily extended to its scaled version, i.e., $(KN{,}N{,}K)$ IC, where we have $d_1(KN{,}N{,}K){=}N{\times}d_1(K{,}1{,}K)$. As it will be clearer later on, $\mathcal{O}_{i^*}(K)$ (i.e., the optimal solution to \eqref{eq:d1_K1K}) refers to the number of co-scheduled transmitters in phase $1$.}
\end{myremark}

A comparison of Theorem \ref{theo:K1K} with the state of the art is shown in Figure \ref{fig:MISO}. As shown, our results yields a significant gain over the two-phase schemes in \cite{Ghasami11,TorrellasK,TorrellasMNK}. Besides, the sum DoF achieved by our scheme is bounded by $\frac{64}{15}$. The proof is shown in the Appendix A. 

\begin{mytheorem}\label{theo:gmat}
\textbf{MAT-like transmission:} For a $(M{,}N{,}K)$ MIMO IC where $1{\leq}\frac{M}{N}{\leq}K$ and $K{\geq}2$, with perfect completely outdated CSIT, considering a $n$-transmitter/$n$-user scheduling in phase $1$ ($2{\leq}n{\leq}K$), an achievable sum DoF is
{\small\begin{IEEEeqnarray}{rcl}
d_1^{mat}(n{,}M{,}N{,}K)&{=}&\tilde{M}n\left(1{+}\frac{\tilde{M}(n{-}1)}{d_2(M{,}N{,}K)}\right)^{-1}{,}\text{\rm almost surely,}\label{eq:d1gmat}
\end{IEEEeqnarray}}
where $\tilde{M}{\triangleq}\min\{M{,}nN\}$, and $d_2(M{,}N{,}K)$ is obtained via Theorem \ref{theo:orderm} by replacing $m{=}2$.
\end{mytheorem}
\begin{mytheorem}\label{theo:RTPIN}
\textbf{Two-stage RT and PIN:} For a $(M{,}N{,}K)$ MIMO IC where $1{\leq}\frac{M}{N}{\leq}K$ and $K{\geq}3$, with perfect completely outdated CSIT, considering a $n$-transmitter/$n$-user scheduling in phase $1$ ($3{\leq}n{\leq}K$), an achievable sum DoF is
{\small\begin{IEEEeqnarray}{rcl}
d_1^{rt{-}pin}(n{,}M{,}N{,}K)&{=}&\hat{M}(n{-}1)\left[\frac{1}{n}\left(n{-}1{+}\frac{\hat{M}}{N(n{-}1)}\right){+}\frac{\hat{M}(n{-}2)}{d_2(M{,}N{,}K)}\right]^{-1}
{,}\text{\rm almost surely,}\label{eq:d1rtpin}
\end{IEEEeqnarray}}
where $\hat{M}{\triangleq}\min\left\{M{,}\frac{1{+}(n{-}1)^2}{1{+}(n{-}2)(n{-}1)}N\right\}$, and $d_2(M{,}N{,}K)$ is obtained via Theorem \ref{theo:orderm} by replacing $m{=}2$.
\end{mytheorem}
\begin{mytheorem}\label{theo:orderm}
For a $(M{,}N{,}K)$ MIMO IC where $1{\leq}\frac{M}{N}{\leq}K$, with perfect completely outdated CSIT, an achievable DoF of delivering order-$m$ messages (for $2{\leq}m{\leq}K$) defined in Definition \ref{def:orderm} is
{\small\begin{IEEEeqnarray}{rcl}
d_m(M{,}N{,}K)&{=}&\left(1{-}A_m(M{,}N{,}K)\right)^{-1},\text{\rm almost surely,}\label{eq:dm}
\end{IEEEeqnarray}}
where
\small{\begin{IEEEeqnarray}{rcl}\label{eq:Am}
\text{\rm For }&\,&2{\leq}m{\leq}K{-}\lceil\frac{M}{N}\rceil{+}1,
A_m(M{,}N{,}K){=}A_{K{-}\lfloor\frac{M}{N}\rfloor{+}1}(M{,}N{,}K)\cdot\Theta_m{+}
\sum_{l{=}m}^{K{-}\lfloor\frac{M}{N}\rfloor}\Delta_{m{,}l},\IEEEyessubnumber\label{eq:Am1}\\
&&\Theta_m{=}\frac{m{-}1}{K{-}\lfloor\frac{M}{N}\rfloor}M^{K{-}\lfloor\frac{M}{N}\rfloor{-}m{+}1}\prod_{i{=}m}^{K{-}\lfloor\frac{M}{N}\rfloor}
\frac{K{-}i}{(M{+}N)(K{-}i){+}N},\IEEEyessubnumber\label{eq:Thetam}\\
&&\Delta_{m{,}l}{=}\frac{M(K{-}l)\left(1{-}\frac{1}{N(l{+}1)}\right){+}l(N{-}1)(K{-}l{+}1)}{l[(M{+}N)(K{-}l){+}N]}\cdot\frac{m{-}1}{l{-}1}M^{l{-}m} \prod_{i{=}m}^{l{-}1}\frac{K{-}i}{(M{+}N)(K{-}i){+}N};\IEEEyessubnumber\label{eq:Deltaml}\\
\text{\rm For }&\,&K{-}\lfloor\frac{M}{N}\rfloor{+}1{\leq}m{\leq}K,A_m(M{,}N{,}K){=}1{-}\frac{1}{N}{+}
\frac{1}{N}\sum_{l{=}m}^{K{-}1}
\frac{(m{-}1)(K{-}l)}{(K{-}m{+}1)(l{-}1)(l{+}1)}.
\IEEEyessubnumber\label{eq:Am2}
\end{IEEEeqnarray}}
\end{mytheorem}

Using the results stated in Theorem \ref{theo:gmat} and \ref{theo:RTPIN}, we obtain a greater achievable sum DoF by taking the maximum of them, which is specified as follows.
\begin{mycoro}\label{coro:sumdof}
For a $(M{,}N{,}K)$ MIMO IC where $1{\leq}\frac{M}{N}{\leq}K$, with perfect completely outdated CSIT, an achievable sum DoF is given by
\begin{IEEEeqnarray}{rcl}
d_1(M{,}N{,}K)&{=}&\max\{\max_{2{\leq}n{\leq}K}d_1^{mat}(n{,}M{,}N{,}K){,}
\max_{3{\leq}n{\leq}K}d_1^{rt{-}pin}(n{,}M{,}N{,}K)\}\nonumber\\
&{=}&\max_{3{\leq}n{\leq}K}\left\{\begin{array}{ll}d_1^{rt{-}pin}(n{,}M{,}N{,}K) & 1{\leq}\frac{M}{N}{\leq}\epsilon(n){;} \\
d_1^{mat}(n{-}1{,}M{,}N{,}K) & \frac{M}{N}{\geq}\epsilon(n){,}\end{array}\right.{,}\text{\rm almost surely,}\label{eq:d1}
\end{IEEEeqnarray}
where $\epsilon(n){=}\frac{n(n{-}1){+}n(n{-}1)^3}{1{+}2(n{-}1)^2{+}(n{-}2)(n{-}1)^3}$.
\end{mycoro}

Note that \eqref{eq:d1} is obtained by 1) the fact that $d_1^{mat}(K{,}M{,}N{,}K){\leq}d_1^{mat}(K{-}1{,}M{,}N{,}K)$, and 2) comparing $d_1^{rt{-}pin}(n{,}M{,}N{,}K)$ with $d_1^{mat}(n{-}1{,}M{,}N{,}K)$ for a fixed $n$, $3{\leq}n{\leq}K$. In Figure \ref{fig:K3} and \ref{fig:K6}, we plot the normalized sum DoF $\frac{d_1(M{,}N{,}K)}{N}$ as a function of the ratio $\rho{\triangleq}\frac{M}{N}$ for $K{=}3$ and $K{=}6$, respectively. Besides, the black dashed curve and the green dotted curve are produced by numerically calculating $\max_{2{\leq}n{\leq}K}d_1^{mat}(n{,}M{,}N{,}K)$ and $\max_{3{\leq}n{\leq}K}d_1^{rt{-}pin}(n{,}M{,}N{,}K)$, respectively. As shown, our result (Corollary \ref{coro:sumdof}, the red solid curve) significantly outperforms the previously known result in \cite{TorrellasMNK} for $M{\geq}N$.
\begin{myremark}
\emph{Theorem \ref{theo:K1K} can be obtained by substituting $M{=}K{,}N{=}1$ into \eqref{eq:d1} and finding the optimal value of $n$ that maximizes \eqref{eq:d1}. Besides, replacing $M{=}N{=}1$ and $n{=}K$ into \eqref{eq:d1rtpin} leads to the sum DoF achieved in \cite{Abdoli13}. Moreover, as shown in Figure \ref{fig:SumDoF11K}, when $4{\leq}K{\leq}13$, our scheme outperforms \cite{Abdoli13} with a Tx-Rx pair scheduling in phase $1$. When $4{\leq}K{\leq}13$, the optimal number of co-scheduled transmitters is $4$. For other values of $K$, scheduling all the Tx-Rx pairs yields the greatest sum DoF performance.}
\end{myremark}

In \cite{Tse10}, the sum DoF of $K$-user MISO BC with delayed CSIT was found with a tight upper-bound. This upper-bound is obtained via a genie-aided model, which gives one user's observation to the others so as to construct physically degraded channels. The genie-aided model was also used in \cite{VVICdelay,xinping_mimo} for two-user MIMO IC. However, in a $K$-user IC, due to the facts that 1) each transmitter only has the access to the message of its related user, and 2) each user overhears multiple interferers, the genie-aided model yields a loose upper-bound. Hence, in this paper, we only focus on the achievable sum DoF.

In the following two sections, we will firstly introduce our proposed scheme for the $(K{,}1{,}K)$ MISO IC, and secondly discuss the generalization in the MIMO case with $1{\leq}\frac{M}{N}{<}K$. 

%% file: ach_K1K.tex
In this section, we focus on the $(K{,}1{,}K)$ IC with perfect outdated CSIT, and study the achievable sum DoF by integrating the MAT-like transmission with the $K$-phase RIA framework proposed in \cite{Abdoli13}.

\subsection{Achievable scheme for the $(3{,}1{,}3)$ IC}\label{sec:ach_313}
According to Theorem \ref{theo:K1K}, when $K{=}3$, one has $\mathcal{O}_1(3){=}2$ and $\mathcal{O}_2(3){=}3$, both leading to the sum DoF $\frac{3}{2}$ according to \eqref{eq:d1_K1K}. This implies that such a sum DoF can be achieved by performing a $3$-transmitter/$3$-user scheduling or a $2$-transmitter/$2$-user scheduling in phase $1$. Let us firstly focus on the case with $3$-transmitter/$3$-user scheduling. The $2$-transmitter/$2$-user approach will be presented afterwards.

The sum DoF $\frac{3}{2}$ is achieved by sending $6$ symbols per user in $12$ slots. The transmission consists of three phases. In phase 1, $6$ symbols per Rx are transmitted in $2$ slots and $12$ order-$2$ symbols are generated. Phase 2 delivers those order-$2$ symbols in $6$ slots, resulting in $3$ order-$3$ symbols and $3$ order-$(1{,}2)$ symbols, which are transmitted using 3 slots in phase 3-I and 1 slot in phase 3-II respectively.
\subsubsection{Phase 1}\label{sec:M3phase1}
\begin{figure}[t]
\renewcommand{\captionfont}{\small}
\captionstyle{center}
\centering
\includegraphics[width=0.65\textwidth,height=3cm]{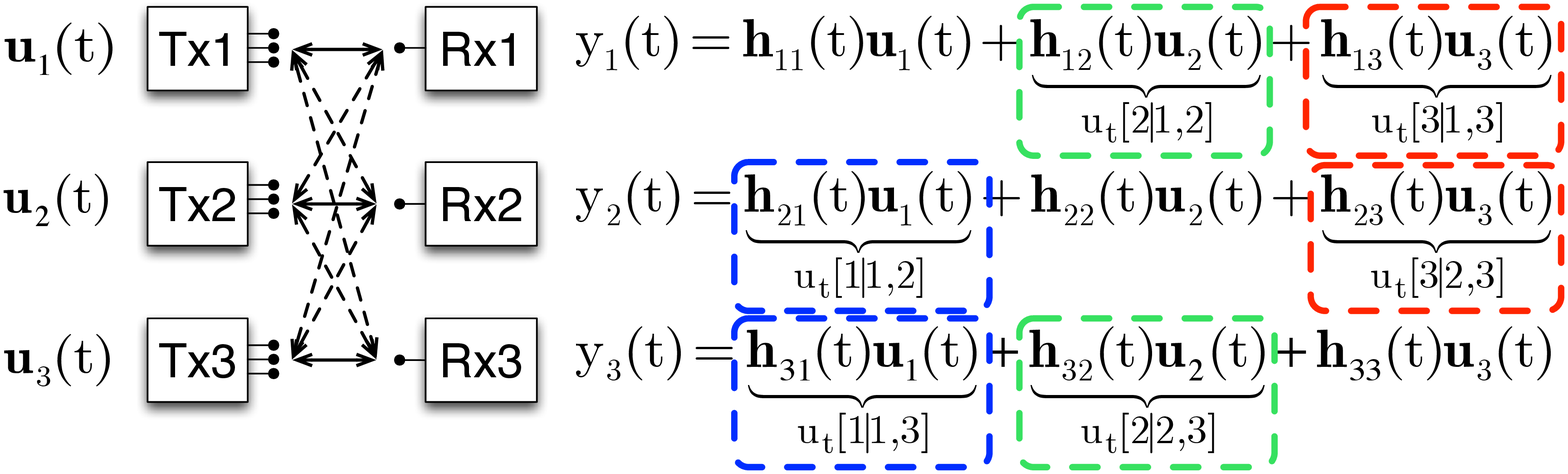}
\caption{Achievable scheme for the $(3{,}1{,}3)$ IC, phase $1$, where $t{=}1{,}2$}\label{fig:phase1_313}
\end{figure}
The transmission lasts for $2$ slots. In each slot, each Tx sends $3$ different symbols to its corresponding user, i.e., $\mathbf{u}_k(t){\in}\mathbb{C}^{3{\times}1}{,}k{=}1{,}2{,}3{,}t{=}1{,}2$. The received signals are illustrated in Figure \ref{fig:phase1_313}, where the noise term is ignored for convenience.

For clarity, let us focus on Rx1, who receives $\mathbf{u}_1(t)$ with other two interferences. Clearly, $\mathbf{u}_1(t)$ can be decoded if 1) $u_t[2|1{,}2]$ and $u_t[3|1{,}3]$ are removed; 2) $u_t[1|1{,}2]$ and $u_t[1|1{,}3]$ are provided to Rx1 in order to have enough linearly independent observations of $\mathbf{u}_1(t)$. Similarly, Rx2 and Rx3 can decode their desired symbols if the interferences are removed and the useful side information is provided. In this way, $u_t[i|i{,}j]{,}i{\neq}j{,}$ is an order-$2$ symbol that is desired by Rx$i$ and Rx$j$. Totally $12$ order-$2$ symbols result from these two slots. The remaining work consists in multicasting $u_1[1|1{,}2]$, $u_2[1|1{,}2]$, $u_1[2|1{,}2]$ and $u_2[2|1{,}2]$ to Rx1 and Rx2, $u_1[1|1{,}3]$, $u_2[1|1{,}3]$, $u_1[3|1{,}3]$ and $u_2[3|1{,}3]$ to Rx1 and Rx3 and $u_1[3|3{,}2]$, $u_2[3|3{,}2]$, $u_1[2|3{,}2]$ and $u_2[2|3{,}2]$ to Rx2 and Rx3.


\subsubsection{Phase 2}\label{sec:K3phase2}
\begin{figure}[t]
\renewcommand{\captionfont}{\small}
\captionstyle{center}
\centering
\includegraphics[width=1\textwidth,height=3.5cm]{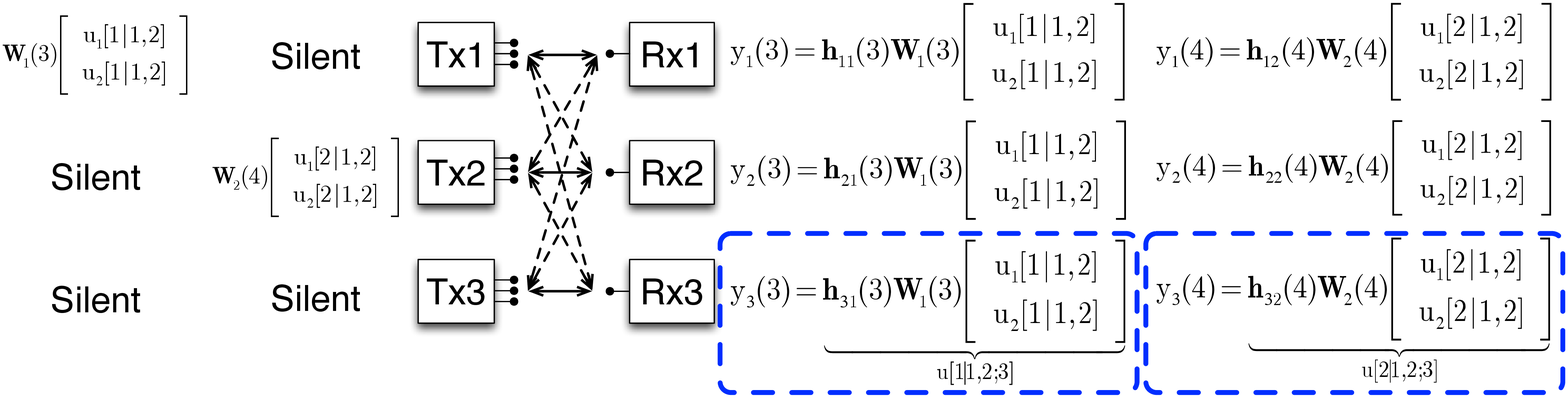}
\caption{Achievable scheme for the $(3{,}1{,}3)$ IC, phase $2$, slot $3$ and $4$}\label{fig:phase2_313}
\end{figure}
We consider that only one transmitter is scheduled per slot, delivering order-$2$ symbols to the corresponding two users, i.e., $1$-Tx/$2$-user scheduling. Specifically, in slot $3$ and $4$, Tx$1$ and Tx$2$ respectively transmit order-$2$ symbols to Rx$1$ and Rx$2$. The transmitted and received signals in these two slots are illustrated in Figure \ref{fig:phase2_313}, where $\mathbf{W}_1(3)$ and $\mathbf{W}_2(4)$ are full rank matrices of size $3{\times}2$. The order-$2$ symbols $u_1[1|1{,}2]$ and $u_2[1|1{,}2]$ (resp. $u_1[2|1{,}2]$ and $u_2[2|1{,}2]$) become decodable at Rx1 and Rx2, if $u[1|1{,}2{;}3]$ (resp. $u[2|1{,}2{;}3]$) is provided to them as such a piece of side information is linear independent of $y_1(3)$ and $y_2(3)$ (resp. $y_1(4)$ and $y_2(4)$).

Similarly, the transmissions in slot 5 to 8 are summarized in Table \ref{tab:phaseIIM3}, where we can see that any two consecutive slots are employed to deliver order-$2$ symbols to a certain subset of two users. To sum up, the transmission is finalized if $u[1|1{,}2{;}3]$ and $u[2|1{,}2{;}3]$ are provided to Rx1 and Rx2, $u[1|1{,}3{;}2]$ and $u[3|1{,}3{;}2]$ are provided to Rx1 and Rx3, while $u[2|2{,}3{;}1]$ and $u[3|2{,}3{;}1]$ are provided to Rx2 and Rx3.
\begin{table}[t]
\captionstyle{center} \centering
\renewcommand{\captionfont}{\small}
\begin{tabular}{c|cccc}
slot & Tx & Rx1 & Rx2 & Rx3\\ \hline
5 & Tx1: $u_1[1{|}1{,}3]{,}u_2[1{|}1{,}3]$ & $y_1(5)$ & $u[1{|}1{,}3;2]$ & $y_3(5)$\\
6 & Tx3: $u_1[3{|}1{,}3]{,}u_2[3{|}1{,}3]$ & $y_1(6)$ & $u[3{|}1{,}3;2]$ & $y_3(6)$\\
7 & Tx2: $u_1[2{|}2{,}3]{,}u_2[2{|}2{,}3]$ & $u[2{|}2{,}3;1]$ & $y_2(7)$ & $y_3(7)$\\
8 & Tx3: $u_1[3{|}2{,}3]{,}u_2[3{|}2{,}3]$ & $u[3{|}2{,}3;1]$ & $y_2(8)$ & $y_3(8)$
\end{tabular}
\caption{The transmission in slot 5 to 8 for the $(3{,}1{,}3)$ IC with $3$-transmitter scheduling in phase $1$.}\label{tab:phaseIIM3}
\end{table}
\subsubsection{Phase 3}\label{sec:K3phase3}
Following the distributed higher order symbol generation proposed in \cite{Abdoli13}, we form order-3 symbols as:
\begin{IEEEeqnarray}{rcl}\label{eq:order3K3}
u[1|1{,}2{,}3]&{=}&LC(u[1|1{,}2{;}3]{,}u[1|1{,}3{;}2]),\IEEEyessubnumber\label{eq:u1_123}\\
u[2|1{,}2{,}3]&{=}&LC(u[2|1{,}2{;}3]{,}u[2|2{,}3{;}1]),\IEEEyessubnumber\label{eq:u2_123}\\
u[3|1{,}2{,}3]&{=}&LC(u[3|1{,}3{;}2]{,}u[3|2{,}3{;}1]),\IEEEyessubnumber\label{eq:u3_123}
\end{IEEEeqnarray}
where $LC$ is short for \emph{Linear Combination}. $u[1|1{,}2{,}3]$, $u[2|1{,}2{,}3]$ and $u[3|1{,}2{,}3]$ are respectively transmitted using a single antenna by Tx1 in slot 9, Tx2 in slot 10 and Tx3 in slot 11 (namely phase 3-I). Consequently, Rx1 is able to decode $u[2|1{,}2{;}3]$ and $u[3|1{,}3{;}2]$ from \eqref{eq:u2_123} and \eqref{eq:u3_123}, respectively, because $u[2|2{,}3{;}1]$ and $u[3|2{,}3{;}1]$ can be removed using the past received signals at Rx1 (see slot $7$ and $8$ shown in Table \ref{tab:phaseIIM3}). However, Rx1 needs one more linear observation to decode $u[1|1{,}2{;}3]$ and $u[1|1{,}3{;}2]$. Similarly, Rx2 (resp. Rx3) needs one more observation to decode $u[2|1{,}2{;}3]$ and $u[2|2{,}3{;}1]$ (resp. $u[3|1{,}3{;}2]$ and $u[3|2{,}3{;}1]$). To this end, in the $12$th slot (phase 3-II), each Tx creates an order-$(1{,}2)$ as
\begin{equation}
u[k|k;i{,}j]{=}LC(u[k|k{,}i{;}j]{,}u[k|k{,}j{;}i]){,}k{\neq}i{\neq}j,
\end{equation}
which is linearly independent of \eqref{eq:order3K3} to prevent from aligning with the observations in phase $3$-I. These three order-$(1{,}2)$ symbols are transmitted simultaneously in slot 12. Rx$1$ is able to obtain an interference-free reception of $u[1{|}1{;}2{,}3]$, because the useful signals contained in $u[2{|}2{;}1{,}3]$ and $u[3{|}3{;}1{,}2]$ have been recovered by Rx$1$ after decoding the order-$3$ symbols. Rx$2$ and Rx$3$ follow similarly. In this way, each user is able to decode the desired signal, so as to proceed to recover order-$2$ and private symbols.

Without the transmission in phase $2$ and $3$, and the generation of the order-$3$ and order-$(1{,}2)$ symbols, the $12$ order-$2$ symbols created in phase $1$ have to be delivered one by one, leading to the requirement of $12$ slots (rather than $10$). Then, the sum DoF would be $\frac{18}{14}$, which is the same as in \cite{Ghasami11}. Besides, in the 2-phase scheme proposed in \cite{TorrellasK}, the new symbol transmission works differently from our scheme. In their scheme, although sending order-$2$ symbols one by one yields the same sum DoF $\frac{3}{2}$ for $K{=}3$, it costs a huge number of time slots when $K$ is large.

\subsubsection{Scheduling $2$ Tx-Rx pairs in phase $1$}
Previous scheme relies on a $3$-transmitter/$3$-user scheduling and requires $3$-transmit antennas in phase 1. Alternatively, we can also use $2$-transmit antenna strategy in phase 1 and employ a $2$-transmitter/$2$-user scheduling. Specifically, we consider that Tx1 and Tx2 are active in slot 1 and 2, Tx1 and Tx3 are active in slot 3 and 4, while Tx2 and Tx3 are active in slot 5 and 6. In each slot, each scheduled transmitter sends two new symbols to the corresponding user. Let us look at slot 1 and 2, where the signals received by Rx1 and Rx2 write as
{\small\begin{IEEEeqnarray}{rcl}
y_1(1){=}\mathbf{h}_{11}(1)\mathbf{W}_1(1)\mathbf{u}_1(1)
{+}\underbrace{\mathbf{h}_{12}(1)\mathbf{W}_2(1)\mathbf{u}_2(1)}_{u_1[2|1{,}2]},&\quad&
y_2(1){=}\underbrace{\mathbf{h}_{21}(1)\mathbf{W}_1(1)\mathbf{u}_1(1)}_{u_1[1|1{,}2]}
{+}\mathbf{h}_{22}(1)\mathbf{W}_2(1)\mathbf{u}_2(1),\IEEEyessubnumber\label{eq:y12t1M2}\\
y_1(2){=}\mathbf{h}_{11}(2)\mathbf{W}_1(2)\mathbf{u}_1(2)
{+}\underbrace{\mathbf{h}_{12}(2)\mathbf{W}_2(2)\mathbf{u}_2(2)}_{u_2[2|1{,}2]},&\quad&
y_2(2){=}\underbrace{\mathbf{h}_{21}(2)\mathbf{W}_1(2)\mathbf{u}_1(2)}_{u_2[1|1{,}2]}
{+}\mathbf{h}_{22}(2)\mathbf{W}_2(2)\mathbf{u}_2(2),\IEEEyessubnumber\label{eq:y12t2M2}
\end{IEEEeqnarray}}where $\mathbf{W}_k(t)$ is a full rank $3{\times}2$ matrix, the symbol vector $\mathbf{u}_k(t)$ is of size $2{\times}1$, for $t{=}1{,}2$ and $k{=}1{,}2$. The received signal $y_3(1)$ and $y_3(2)$ are not shown as Tx$3$ and Rx$3$ are silent in slot 1 and 2. We can see that both user 1 and user 2 can decode their desired symbols if $u_1[1{|}1{,}2]$ and $u_2[1{|}1{,}2]$ are exchanged with $u_1[2{|}1{,}2]$ and $u_2[2{|}1{,}2]$. Similar transmissions are performed in slot 3, 4, 5 and 6. Thus, totally $24$ new symbols (e.g. $8$ per Rx) are sent in $6$ slots, generating $12$ order-$2$ symbols.

Applying the higher order symbol transmission introduced in Section \ref{sec:K3phase2} and \ref{sec:K3phase3}, those $12$ order-$2$ symbols are successfully delivered in $10$ slots, yielding the sum DoF $\frac{24}{16}{=}\frac{3}{2}$. Clearly, this scheme is applicable to the case where each transmitter is equipped with $2$ antennas, as the transmissions in all the three phases rely on at most $2$ transmit antennas.

Next, we present the general transmission strategy for the $(K{,}1{,}K)$ IC.

\subsection{Generalized Scheme for $(K{,}1{,}K)$ MISO IC}\label{sec:scheme}
\begin{figure}[t]
\renewcommand{\captionfont}{\small}
\captionstyle{center}
\centering
\includegraphics[width=0.8\textwidth,height=2.25cm]{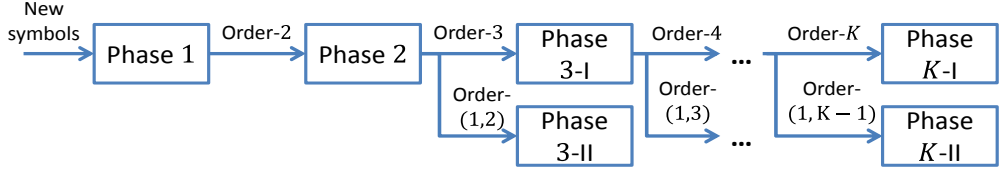}
\caption{Transmission flow}\label{fig:flow}
\end{figure}
\subsubsection{Transmission and Decoding Flow}\label{sec:flow}
Similar to \cite{Abdoli13}, the $K$-phase transmission is illustrated in Figure \ref{fig:flow}. All the private symbols are transmitted in phase 1, generating order-2 symbols. Then, all the order-2 symbols are delivered in phase 2. At the end of phase 2, two types of higher order symbols are generated, namely order-3 and order-(1,2) symbols, which will be delivered in phase 3-I and 3-II respectively. This transmission is repeated till phase $K$, where order-$K$ and order-$(1{,}K{-}1)$ symbols are delivered.

A backward decoding is carried out. Specifically, each user recovers order-$K$ and $(1{,}K{-}1)$ symbols first. Then with their knowledge, order-$(K{-}1)$ symbols can be decoded. Repeatedly, order-$m$ symbols ($m{\geq}2$) are recovered using order-$(m{+}1)$ and $(1{,}m)$ symbols. At last, all the private symbols are decoded with the knowledge of order-$2$ symbols.

In this way, considering that $N_1$ private symbols are sent in $T_1$ slots in phase 1, generating $N_2$ order-$2$ symbols, the achievable sum DoF can be computed as
\begin{IEEEeqnarray}{rcl}
d_1(K{,}1{,}K)&{=}&\frac{N_1}{T_1{+}\frac{N_2}{d_2(K{,}1{,}K)}}.\label{eq:d1_K1K_recur}
\end{IEEEeqnarray}
Following the aforementioned transmission flow, the DoF of delivering order-$2$ symbols, i.e., $d_2(M{,}N{,}K)$, can be computed recursively as, for $2{\leq}m{\leq}K{-}1$,
\begin{IEEEeqnarray}{rcl}
d_m(K{,}1{,}K)&{=}&\frac{N_m}{T_m{+}\frac{N_{m{+}1}}{d_{m{+}1}(K{,}1{,}K)}{+}\frac{N_{1{,}m}}{d_{1{,}m}(K{,}1{,}K)}},\label{eq:dofm_K1K_recur}
\end{IEEEeqnarray}
where $N_m$ and $N_{1{,}m}$ represent the number of order-$m$ and $(1{,}m)$ symbols respectively. $d_m(K{,}1{,}K)$ and $d_{1{,}m}(K{,}1{,}K)$ stand for the DoF of sending order-$m$ and $(1{,}m)$ symbols, respectively. $T_m$ refers to the number of slots in phase $m$-I. Note that we have $d_K(K{,}1{,}K){=}1$ due to the fact that order-$K$ symbols are intended for all users and each user is equipped with a single antenna. Besides, $d_{1{,}m}(K{,}1{,}K){=}m{+}1$ will be shown in Section \ref{sec:mp1K1K}. Next, the work is reduced to quantify 1) $N_1$, $N_2$ and $T_1$ in \eqref{eq:d1_K1K_recur}, and 2) the parameters $N_m$, and $N_{m{+}1}$, $T_m$ and $N_{1{,}m}$ for $2{\leq}m{\leq}K{-}1$ in \eqref{eq:dofm_K1K_recur}.

\subsubsection{Phase 1}\label{sec:phase1_K1K}
We consider an $n$-transmitter/$n$-user scheduling. Specifically, in a certain slot, a subset of $\mathcal{S}_n$ ($n{\leq}K$ to be shown later on) transmitters are active while others keep silent. Each of them delivers $n$ new symbols to the corresponding user, i.e., $\mathbf{u}_k{\in}\mathbb{C}^{n{\times}1}{,}{\forall}k{\in}\mathcal{S}_n$. The precoder $\mathbf{W}_k$ used by Tx$k$ is a full rank matrix of size $K{\times}n$.

The received signal writes as $y_k{=}\sum_{j{\in}\mathcal{S}_n}\mathbf{h}_{kj}\mathbf{W}_j\mathbf{u}_j$. We can see that if any two scheduled users, Rx$k$ and Rx$j$, ${\forall}k{,}j{\in}\mathcal{S}_n{,}k{\neq}j$, exchange their side information, i.e., $\mathbf{h}_{kj}\mathbf{W}_j\mathbf{u}_j$ and $\mathbf{h}_{jk}\mathbf{W}_k\mathbf{u}_k$, then each user obtains $n$ interference-free linear observations of its desired symbols, i.e., $\{\mathbf{h}_{jk}\mathbf{W}_k\mathbf{u}_k\}_{{\forall}j{\in}\mathcal{S}_n}$. These $n$ linear observations are linearly independent of each other since channels are i.i.d. across the users and the number of transmit antennas are large enough. Therefore, the term $\mathbf{h}_{kj}\mathbf{W}_j\mathbf{u}_j$ is an order-$2$ symbol as it is useful for Rx$k$ for interference cancelation/alignment, and for Rx$j$ as a useful side information.

It is straightforward that $n(n{-}1)$ order-$2$ symbols (e.g. $n$ receivers and each with $n{-}1$ interferers) are generated in a certain slot. Besides, since there are ${{K}\choose{n}}$ possible choices of $\mathcal{S}_n$, the same transmission is repeated ${{K}\choose{n}}$ times for transmitter scheduling. Thus, one has
\begin{IEEEeqnarray}{rcl}
N_1{=}n^2{{K}\choose{n}},\quad&
T_1{=}{{K}\choose{n}},\quad&
N_2{=}n(n{-}1){{K}\choose{n}}.\label{eq:N1T1N2_K1K}
\end{IEEEeqnarray}
Then, the sum DoF with $n$-transmitter/$n$-user scheduling in phase $1$, i.e., $d_1(n{,}K{,}1{,}K)$, is written as
\begin{equation}
d_1(n{,}K{,}1{,}K){=}\frac{n^2}{1{+}\frac{n(n{-}1)}{d_2(K{,}1{,}K)}}.\label{eq:d1_K1K_recur2}
\end{equation}
Hence, the optimal $n$ is chosen such that,
\begin{equation}
n^*{=}arg\max_{n{=}2{,}\cdots{,}K}\frac{n^2}{1{+}\frac{n(n{-}1)}{d_2(K{,}1{,}K)}}.\label{eq:nphase1}
\end{equation}
By evaluating the first and second order derivatives of \eqref{eq:d1_K1K_recur2}, one can easily find that the global minimizer is given by $2d_2(K{,}1{,}K)$. As $n^*$ is an integer, we choose $n^*$ to be either $\lfloor2d_2(K{,}1{,}K)\rfloor$ or $\lceil2d_2(K{,}1{,}K)\rceil$. This leads to the maximization operator in \eqref{eq:d1_K1K}. The remaining work is to find $d_2(K{,}1{,}K)$.

\subsubsection{Phase $m$-I ($2{\leq}m{\leq}K{-}1$)}\label{sec:phasem}
We perform a $1$-Tx/$m$-user scheduling and employ the same transmission strategy in MAT. To be specific, in a certain slot and for a subset $\mathcal{S}_m$ of $m$ users, only one transmitter, i.e.Tx$k{,}k{\in}\mathcal{S}_m$, is active, delivering $K{-}m{+}1$ order-$m$ symbols, i.e., $\mathbf{u}[k|\mathcal{S}_m]{\in}\mathbb{C}^{K{-}m{+}1}$. The precoder $\mathbf{W}_k$ used by Tx$k$ is a full rank  $K{\times}(K{-}m{+}1)$ matrix.

If the non-scheduled $K{-}m$ users provide their received signals to the $m$ scheduled users, each scheduled user obtains $K{-}m{+}1$ linear observations of the desired order-$m$ symbols. For a Rx$j{,}j{\in}\mathcal{S}_m$, let us write the $K{-}m{+}1$ linear observations as $y_j{,}\{y_l\}_{{\forall}l{\notin}\mathcal{S}_m}$. Note that these $K{-}m{+}1$ are linearly independent of each other since the channels are i.i.d. across the users and the number of transmit antennas are large enough. Then, Rx$j{,}{\forall}j{\in}\mathcal{S}_m$ will be able to decode $\mathbf{u}[k|\mathcal{S}_m]$. Therefore, we denote the received signals at the $K{-}m$ non-scheduled users as $u[k|\mathcal{S}_m;l]{,}{\forall}l{\notin}\mathcal{S}_m$, as they are useful side information for the $m$ scheduled uses of $\mathcal{S}_m$.

As there are $m$ choices of $k$ in $\mathcal{S}_m$ and there are ${{K}\choose{m}}$ possible choices of $\mathcal{S}_m$, we have
\begin{IEEEeqnarray}{rcl}
T_m{=}m{{K}\choose{m}},&\quad& N_m{=}(K{-}m{+}1)T_m{,}\label{eq:NmTm_K1K}
\end{IEEEeqnarray}
and the total number of useful signals is $(K{-}m)T_m$. Next, we employ these useful signals to formulate order-$(m{+}1)$ and order-$(1{,}m)$ symbols.

\begin{table}[t]
\captionstyle{center} \centering
\renewcommand{\captionfont}{\small}
\begin{tabular}{c|c|c|c|c|c}
 & Tx$1$ & Tx$2$ & Tx$3$ & $\cdots$ & Tx$m{+}1$\\ \hline
Rx$1$ & & $u[2{|}\mathcal{S}_{m{+}1}{\setminus}1{;}1]$ & $u[3{|}\mathcal{S}_{m{+}1}{\setminus}1{;}1]$ & $\cdots$ & $u[m{+}1{|}\mathcal{S}_{m{+}1}{\setminus}1{;}1]$\\ \hline
Rx$2$ & $u[1{|}\mathcal{S}_{m{+}1}{\setminus}2{;}2]$ & & $u[3{|}\mathcal{S}_{m{+}1}{\setminus}2{;}2]$ & $\cdots$ & $u[m{+}1{|}\mathcal{S}_{m{+}1}{\setminus}2{;}2]$\\ \hline
$\vdots$ & $\vdots$ & $\vdots$ & $\vdots$ & $\vdots$ & $\vdots$ \\ \hline
Rx$m{+}1$ & $u[1{|}\mathcal{S}_{m{+}1}{\setminus}m{+}1{;}m{+}1]$ & $u[2{|}\mathcal{S}_{m{+}1}{\setminus}m{+}1{;}m{+}1]$ & $u[3{|}\mathcal{S}_{m{+}1}{\setminus}m{+}1{;}m{+}1]$ & $\cdots$ &
\end{tabular}
\caption{The useful side information generated after phase $m$-I ($2{\leq}m{\leq}K{-}1$) in the $(K{,}1{,}K)$ IC.}\label{tab:ordermp1}
\end{table}

To understand the distributed higher order symbol generation, without loss of generality, we present the useful signals obtained by a certain subset $\mathcal{S}_{m{+}1}{=}\{1{,}\cdots{,}m{+}1\}$ of users in Table \ref{tab:ordermp1}. Each row shows the useful signals obtained by a certain user after phase $m$-I, while each column summarizes the useful signals that can be reconstructed by a certain Tx after phase $m$-I. We can see that, for the $m$ useful signals available at Tx$1$, i.e., $u[1{|}\mathcal{S}_{m{+}1}{\setminus}j{;}j]{,}{\forall}j{=}2{,}\cdots{,}m{+}1$, Rx$1$ wishes to decode all of them, while Rx$j{,}j{=}2{,}\cdots{,}m{+}1$ wishes to decode $m{-}1$ useful signals (except $u[1{|}\mathcal{S}_{m{+}1}{\setminus}j{;}j]$).

To this end, using the $m$ useful signals, Tx$1$, firstly formulates $m{-}1$ linear combinations. These $m{-}1$ linear combinations are desired by all the $m{+}1$ users and known as order-$(m{+}1)$ symbols, i.e., $\mathbf{u}[1{|}\mathcal{S}_{m{+}1}]{\in}\mathbb{C}^{(m{-}1){\times}1}$. Once they are delivered to all the users in $\mathcal{S}_{m{+}1}$, Rx$j$, $j{=}2{,}\cdots{,}m{+}1$ is able to decode the desired $m{-}1$ useful signals $\mathbf{u}[1{|}\mathcal{S}_{m{+}1}{\setminus}l{;}l]{,}l{\neq}j{,}l{,}j{=}2{,}\cdots{,}m{+}1$. However, Rx$1$ does not have plenty of linear observations to decode all the desired $m$ useful signals. Toward this, Tx$1$ formulates another linear independent linear combination, which is known as order-$(1{,}m)$ symbols, i.e., $\mathbf{u}[1{|}1{;}\mathcal{S}_{m{+}1}{\setminus}1]$, because it is desired by only one user, i.e., Rx$1$, and already obtained by the other $m$ users after the order-$(m{+}1)$ symbols are successfully delivered.

Since there are ${{K}\choose{m{+}1}}$ choices of $\mathcal{S}_{m{+}1}$ and $m{+}1$ different transmitters in each $\mathcal{S}_{m{+}1}$, the total number of order-$(m{+}1)$ and order-$(1{,}m)$ symbols are respectively given by
\begin{IEEEeqnarray}{rcl}
N_{m{+}1}{=}(m{-}1)(m{+}1){{K}\choose{m{+}1}},&\quad&
N_{1{,}m}{=}(m{+}1){{K}\choose{m{+}1}}.\label{eq:Nmp1N1m_K1K}
\end{IEEEeqnarray}

\subsubsection{Phase $(m{+}1)$-II, for $2{\leq}m{\leq}K{-}1$}\label{sec:mp1K1K}
In this subphase, order-$(1{,}m)$ symbols are transmitted. The transmission strategy is the same as that designed for the $(1{,}1{,}K)$ case \cite{Abdoli13}. Specifically, as an order-$(1{,}m)$ symbol is generated for a subset of $\mathcal{S}_{m{+}1}$ users, the transmission in phase $(m{+}1)$-II is divided into ${{K}\choose{m{+}1}}$ slots. In each slot, a certain subset of $\mathcal{S}_{m{+}1}$ transmitters are active and delivering $m{+}1$ order-$(1{,}m)$ symbols. Let us consider $\mathcal{S}_{m{+}1}{=}\{1{,}\cdots{,}m{+}1\}$ without loss of generality. According to Definition \ref{def:order1m}, Rx$1$ is able to obtain an interference-free reception of $u[1{|}1{;}\mathcal{S}_{m{+}1}{\setminus}1]$, because the order-$(1{,}m)$ symbols sent by other transmitters, i.e., $u[j{|}j{;}\mathcal{S}_{m{+}1}{\setminus}j]{,}j{=}2{,}\cdots{,}m{+}1$ can be removed by Rx$1$ after decoding the order-$(m{+}1)$ symbols $\mathbf{u}[j{|}\mathcal{S}_{m{+}1}]{\in}\mathbb{C}^{(m{-}1){\times}1}{,}j{=}2{,}\cdots{,}m{+}1$ (note that we employ a backward decoding). Then, the achievable DoF of delivering order-$(1{,}m)$ symbols is $d_{1{,}m}(K{,}1{,}K){=}(m{+}1)$.

\subsubsection{Phase $K$-I}
The transmission in phase $K$-I lasts for $K$ slots, where in each slot a certain Tx transmits $1$ order-$K$ symbol, i.e., $u[k|\mathcal{S}_K]$. Since each user has a single antenna, the order-$K$ symbol can be decoded. Then, the achievable DoF of delivering order-$K$ symbols in the $(K{,}1{,}K)$ IC is $d_K(K{,}1{,}K){=}1$.

Plugging \eqref{eq:NmTm_K1K} and \eqref{eq:Nmp1N1m_K1K} into \eqref{eq:dofm_K1K_recur}, the recursive expression of the DoF of delivering order-$m$ symbols, i.e., $d_m(K{,}1{,}K)$, is given by
\begin{IEEEeqnarray}{rcl}
d_m(K{,}1{,}K)&{=}&\frac{m(K{-}m{+}1)}{m{+}\frac{K{-}m}{m{+}1}{+}\frac{(m{-}1)(K{-}m)}{d_{m{+}1}(K{,}1{,}K)}}.\label{eq:dofm_K1K_recur2}
\end{IEEEeqnarray}
Then, $d_2(K{,}1{,}K)$ in \eqref{eq:d2_K1K} holds following the general proof in Appendix B. Combining with the optimization problem \eqref{eq:d1_K1K_recur2} leads to Theorem \ref{theo:K1K}.

So far, we have characterized an achievable sum DoF of a $(K{,}1{,}K)$ IC by integrating MAT-like transmission and the $K$-phase RIA approach proposed in \cite{Abdoli13}. In the next section, we will draw our attention to the general $(M{,}N{,}K)$ IC with $1{\leq}\frac{M}{N}{\leq}K$. Since the $(K{,}1{,}K)$ IC and the $(1{,}1{,}K)$ IC can be regarded as special cases with $M{=}K{,}N{=}1$ and $M{=}1{,}N{=}1$, respectively, we aim at generalizing the achievable schemes in these two scenarios to the general case with $1{\leq}\frac{M}{N}{\leq}K$, or ideally, finding a scheme that smoothly connects them when $M{=}K{,}N{=}1$ and $M{=}1{,}N{=}1$.

%% file: ach_MNK.tex
In this section, we propose two schemes for the $(M{,}N{,}K)$ IC with $1{\leq}\frac{M}{N}{\leq}K$ based on the $K$-phase RIA framework. In phase $1$, these two schemes generalize the MAT-like transmission and the RT-PIN approach proposed in \cite{Abdoli13}. In phase $m$, $2{\leq}m{\leq}K$, the two schemes employ an identical transmission strategy, which originates from the RT-PIN approach proposed in \cite{Abdoli13} and becomes the MAT-like transmission when $M{\geq}N(K{-}m{+}1)$. We will start with a $(3{,}2{,}3)$ IC example and then go into the general case.

\subsection{Achievable schemes for the $(3{,}2{,}3)$ IC}\label{sec:ach323}
\input{ach_323}

\subsection{$(M{,}N{,}K)$ MIMO IC: Phase $1$, MAT-like Transmission}\label{sec:mat}
In this subsection, we show the achievability of MAT-like transmission in the general $(M{,}N{,}K)$ IC with $1{\leq}\frac{M}{N}{\leq}K$. We focus on a $n$-transmitter/$n$-user scheduling in phase $1$ ($2{\leq}n{\leq}K$). Here, we only consider the case $\frac{M}{N}{\leq}n$, because the achievability in the case $\frac{M}{N}{\geq}n$ follows similarly by switching off the redundant transmit antennas. Motivated by the $(3{,}2{,}3)$ IC, we learn that when the number of transmit antennas is insufficiently large, i.e., $M{<}nN$, the signal received by the desired user is linearly dependent of the side information obtained by the other $n{-}1$ scheduled users. Hence, only part of the overheard interferences can be considered as order-$2$ symbols.

Accordingly, we consider that the $n$ co-scheduled transmitters are active for $n$ slots, during which each of them delivers $nM$ new symbols to the corresponding user, i.e., $\mathbf{u}_k{\in}\mathbb{C}^{Mn{\times}1}{,}{\forall}k{\in}\mathcal{S}_n$, and the precoder used by Tx$k$ across the time slots, $\mathbf{W}_k$, is a full rank matrix of size $nM{\times}nM$. At the receiver side, each user randomly obtains $M$ linear observations from the $nN$-dimensional received signal as $\mathbf{P}_k\mathbf{y}_k$ where $\mathbf{P}_k$ is a $M{\times}Nn$ full rank matrix. Then, we see that each user is able to decode their desired signal, if any two scheduled users Rx$k$ and Rx$j$, ${\forall}k{,}j{\in}\mathcal{S}_n$, exchange their side information, i.e., $\mathbf{P}_k\bar{\mathbf{H}}_{kj}\mathbf{W}_j\mathbf{u}_j$ and $\mathbf{P}_j\bar{\mathbf{H}}_{jk}\mathbf{W}_k\mathbf{u}_k$, where $\bar{\mathbf{H}}_{kj}{\triangleq}\{\mathbf{H}_{kj}(t)\}_{t{=}1{,}\cdots{,}n}$ and $\bar{\mathbf{H}}_{jk}{\triangleq}\{\mathbf{H}_{jk}(t)\}_{t{=}1{,}\cdots{,}n}$. Therefore, the $M$ elements of $\mathbf{P}_k\bar{\mathbf{H}}_{kj}\mathbf{W}_j\mathbf{u}_j{\in}\mathbb{C}^{M{\times}1}$ are order-$2$ symbol desired by Rx$k$ and Rx$j$.

Since there are $Mn(n{-}1)$ order-$2$ symbols generated for a certain subset of $n$ users, and there are ${K}\choose{n}$ possible choices of $\mathcal{S}_n$, we have
\begin{IEEEeqnarray}{rcl}
N_1{=}Mn^2{{K}\choose{n}}{,}\quad&T_1{=}n{{K}\choose{n}}{,}\quad&N_2{=}M(n{-}1)n{{K}\choose{n}}{.}\label{eq:N1T1_MAT}
\end{IEEEeqnarray}
As the achievable sum DoF can be expressed as $d_1^{mat}(n{,}M{,}N{,}K){=}\frac{N_1}{T_1{+}N_2/d_2(M{,}N{,}K)}$, Theorem \ref{theo:gmat} holds with the parameters in \eqref{eq:N1T1_MAT}.

\subsection{$(M{,}N{,}K)$ MIMO IC: Phase $1$, RT and PIN}\label{sec:rtpin}
\input{RTPIN}
\subsection{$(M{,}N{,}K)$ MIMO IC: Phase $m$, $2{\leq}m{\leq}K$}\label{sec:phasemMNK}
\input{phasemMNK}

\subsection{Discussions}
\subsubsection{Connection between the schemes}\label{sec:cnt}
Let us focus on RT-PIN approach with $n$ active transmitters in phase $1$ and MAT-like transmission with $(n{-}1)$ active transmitters in phase $1$, where $3{\leq}n{\leq}K$. To link these two schemes, we introduce a parameter, $r{\triangleq}\frac{T_1}{N_2}$, i.e., \emph{the number of time slots needed to generate an order-$2$ symbol}. With $T_1$ and $N_2$ given in \eqref{eq:N1T1_MAT}, \eqref{eq:N1T1_RTPIN} and \eqref{eq:N2_RTPIN}, the ratios write as,
\begin{IEEEeqnarray}{rcl}
r^{mat}(n{-}1{,}M{,}N{,}K){=}\frac{1}{\tilde{M}(n{-}2)}{,}&\quad&
r^{rt{-}pin}(n{,}M{,}N{,}K){=}\frac{(n{-}1)^2N{+}\hat{M}}{n(n{-}1)(n{-}2)\hat{M}N}{,}\quad3{\leq}n{\leq}K{,}\label{eq:ratio}
\end{IEEEeqnarray}
with $\tilde{M}$ in \eqref{eq:d1gmat} and $\hat{M}$ in \eqref{eq:d1rtpin}. Then, the sum DoF is interpreted as $d_1{\triangleq}\frac{N_1/N_2}{r{+}d_2^{-1}}$, namely
\begin{IEEEeqnarray}{rcl}
d_1^{mat}(n{-}1{,}M{,}N{,}K){=}\frac{(n{-}1)/(n{-}2)}{r^{mat}{+}d_2^{-1}}{,}&\quad&
d_1^{rt{-}pin}(n{,}M{,}N{,}K){=}\frac{(n{-}1)/(n{-}2)}{r^{rt{-}pin}{+}d_2^{-1}}{,}\quad3{\leq}n{\leq}K{,}\label{eq:d1r}
\end{IEEEeqnarray}
where the parameters involved in $r^{mat}$, $r^{rt{-}pin}$ and $d_2$ are ignored for convenience. Hence, it is clearly that the ratios in \eqref{eq:ratio} act as indicators showing which scheme yields the better sum DoF performance. By comparing the ratios in \eqref{eq:ratio}, we can reach the concise expression in \eqref{eq:d1}.
\subsubsection{Tx-Rx pairs scheduling in phase $1$}\label{sec:scheduling}
The discussion in Section \ref{sec:cnt} is only useful in judging the proposed schemes for a fixed value of $n$, but cannot be employed to find the optimal solution to Corollary \ref{coro:sumdof}. This is because there exists another important parameter, $\frac{N_1}{N_2}{\triangleq}\frac{n{-}1}{n{-}2}$, that impacts the sum DoF (see \eqref{eq:d1r}). This parameter tells us how many private symbols can be decoded once a single order-$2$ symbol is provided. In general, a greater value of $n$ yields a smaller $r^{mat}$ or $r^{rt{-}pin}$, but leads to a smaller $\frac{n{-}1}{n{-}2}$. Consequently, solving such a trade-off is essential to the sum DoF performance, thus leading to the emergence of performing a proper Tx-Rx pair scheduling in phase $1$.

\subsubsection{Global CSIT vs. Local CSIT}
In phase 1, we can see that the construction of the order-$2$ symbols in the MAT-like transmission (presented in Section \ref{sec:mat}) only relies on \emph{local} CSIT, as $\mathbf{P}_k$ is a random matrix and $\mathbf{G}_{kj}$ is the outgoing channel of Tx$j$. However, in the RT-PIN approach, since the matrix $\mathbf{Q}_{1j}$ in \eqref{eq:Qi1ij} is related to channel matrix $\mathbf{G}_{1j}^{is}$, constructing order-$2$ symbols, e.g., $\bar{\mathbf{G}}_{1l{,}j}\mathbf{u}_l{,}l{\neq}j$ in \eqref{eq:yeffectcase2ph1}, needs \emph{global} CSIT, i.e., the outgoing channels of both Tx$j$ and Tx$l$.

In phase $m$, $2{\leq}m{\leq}K$, In \eqref{eq:PINphasemI}, we can see that the useful side information $\mathbf{u}[1|\mathcal{S}_m;j]$ is obtained by PIN and will be reconstructed by Tx$1$ at the end of phase $m$-I. Since $\mathbf{F}_{j2}$ is related to $\mathbf{G}_{j2}$, namely the outgoing channel of Tx$2$, reconstructing $\mathbf{u}[1|\mathcal{S}_m;j]$ requires \emph{global} CSIT. However, when $M{\geq}N(K{-}m{+}1)$, as there is only one active Tx, it is unnecessary to perform PIN so that only \emph{local} CSIT is needed.

Moreover, we can conclude that, when $\frac{M}{N}{\geq}K{-}1$, the sum DoF stated in Corollary \ref{coro:sumdof} is achievable with local CSIT. The reasons are two-fold. Firstly, when $\frac{M}{N}{\geq}K{-}1$, one can verify that performing MAT-like transmission in phase $1$ yields a greater sum DoF than performing RT-PIN. Secondly, when $\frac{M}{N}{\geq}K{-}1$, only local CSIT is needed to support the transmissions in phase $2$ through to phase $K$.


%% file: ach_323.tex
In this subsection, we aim to show that $d_1^{mat}(2{,}3{,}2{,}3){=}\frac{21}{8}$ and $d_1^{rt{-}pin}(3{,}3{,}2{,}3){=}\frac{504}{185}$, implying that in the $(3{,}2{,}3)$ IC, RT-PIN scheme with $3$-transmitter/$3$-user scheduling in phase $1$ outperforms MAT-like transmission with $2$-transmitter/$2$-user scheduling in phase $1$.
\subsubsection{Phase $1$, MAT-like transmission}
In MAT scheme, the overheard interferences obtained by various users are directly regarded as order-$2$ symbols. However, when the number of transmit antennas at each transmitter is smaller than the total number of received antennas at all the scheduled users i.e., $3{<}2{\times}2$ in the $(3{,}2{,}3)$ IC with $2$ co-scheduled transmitters, the overheard interferences at various users are linearly dependent of the side information obtained by the desired user. To counter this problem, we propose a MAT-like scheme by giving up some overheard interferences.

Here, we consider a 2-transmitter/2-user scheduling. In slot 1 and 2, Tx1 and Tx2 are co-scheduled each delivering $6$ symbols to its corresponding user, and $6$ order-$2$ symbols, $\mathbf{u}_k{\in}\mathbb{C}^{6{\times}1}{,}k{=}1{,}2$, are generated. Specifically, the aggregate transmitted signals write as
\begin{IEEEeqnarray}{rcl}
\mathbf{s}_1{=}{\rm stack}\left\{\mathbf{s}_1(1){,}\mathbf{s}_1(2)\right\}{=}\mathbf{W}_1\mathbf{u}_1{,}&\quad&
\mathbf{s}_2{=}{\rm stack}\left\{\mathbf{s}_2(1){,}\mathbf{s}_2(2)\right\}{=}\mathbf{W}_2\mathbf{u}_2{,}
\end{IEEEeqnarray}
where $\mathbf{W}_k{\in}\mathbb{C}^{6{\times}6}{,}k{=}1{,}2$ is a full rank precoders across the two slots. The received signals write as
\begin{IEEEeqnarray}{rcl}
\mathbf{y}_1{=}
\underbrace{\bar{\mathbf{H}}_{11}\mathbf{W}_1}_{\mathbf{G}_{11}{\in}\mathbb{C}^{4{\times}6}}\mathbf{u}_1{+}
\underbrace{\bar{\mathbf{H}}_{12}\mathbf{W}_2}_{\mathbf{G}_{12}{\in}\mathbb{C}^{4{\times}6}}\mathbf{u}_2{,}&\quad&
\mathbf{y}_2{=}
\underbrace{\bar{\mathbf{H}}_{21}\mathbf{W}_1}_{\mathbf{G}_{21}{\in}\mathbb{C}^{4{\times}6}}\mathbf{u}_1{+}
\underbrace{\bar{\mathbf{H}}_{22}\mathbf{W}_2}_{\mathbf{G}_{22}{\in}\mathbb{C}^{4{\times}6}}\mathbf{u}_2{,}\label{eq:y_323MATphase1}
\end{IEEEeqnarray}
where $\mathbf{y}_k{=}{\rm stack}\left\{\mathbf{y}_k(1){,}\mathbf{y}_k(2)\right\}$ and $\bar{\mathbf{H}}_{kj}{\triangleq}\text{\rm Bdiag}\left\{\mathbf{H}_{kj}(1){,}\mathbf{H}_{kj}(2)\right\}{\in}\mathbb{C}^{4{\times}6}{,}k{,}j{=}1{,}2$ refers to the aggregate channel matrix across the two slots. The effective channel matrix $\bar{\mathbf{H}}_{kj}\mathbf{W}_j$ is denoted by $\mathbf{G}_{kj}$.

At this moment, if we simply exchange $\mathbf{G}_{12}\mathbf{u}_2$ and $\mathbf{G}_{21}\mathbf{u}_1$, each user has totally $8$ linear observations of their $6$ symbols. This implies that there are $2$ redundant observations for each user. Hence, it is improper to treat all the $4$ elements in $\mathbf{G}_{12}\mathbf{u}_2$ (resp. $\mathbf{G}_{21}\mathbf{u}_1$) as order-$2$ symbols. Due to this fact, each user randomly obtains $3$ linear observations from its $4$-dimensional received signal as
\begin{IEEEeqnarray}{rcl}
\bar{\mathbf{y}}_1{=}\mathbf{P}_1\mathbf{y}_1{=}\mathbf{P}_1\mathbf{G}_{11}\mathbf{u}_1{+}
\underbrace{\mathbf{P}_1\mathbf{G}_{12}\mathbf{u}_2}_{\mathbf{u}[2{|}1{,}2]{\in}\mathbb{C}^{3{\times}1}}{,}&\quad&
\bar{\mathbf{y}}_2{=}\mathbf{P}_2\mathbf{y}_2{=}
\underbrace{\mathbf{P}_2\mathbf{G}_{21}\mathbf{u}_1}_{\mathbf{u}[1{|}1{,}2]{\in}\mathbb{C}^{3{\times}1}}{+}
\mathbf{P}_2\mathbf{G}_{22}\mathbf{u}_2{,}\label{eq:ybar_323MATphase1}
\end{IEEEeqnarray}
where $\mathbf{P}_k{\in}\mathbb{C}^{3{\times}4}{,}k{=}1{,}2$ is a full rank matrix. Then, if $\mathbf{P}_1\mathbf{G}_{12}\mathbf{u}_2$ and $\mathbf{P}_2\mathbf{G}_{21}\mathbf{u}_1$ are exchanged, the desired symbols become decodable, because each user obtains a $6{\times}6$ full rank effective channel matrix ${\rm stack}\{\mathbf{P}_k\mathbf{G}_{kk}{,}\mathbf{P}_j\mathbf{G}_{jk}\}{,}k{,}j{=}1{,}2$, almost surely. In this way, the $3$ elements in $\mathbf{P}_k\mathbf{G}_{kj}\mathbf{u}_j{,}k{,}j{=}1{,}2$ are order-$2$ symbols as they are useful signals to Rx$1$ and Rx$2$.

The transmissions in slot 3 and 4 (where Tx1 and Tx3 are active), and the transmissions in slot 5 and 6 (where Tx2 and Tx3 are active) follow similarly. Consequently, we deliver totally $36$ symbols in $6$ slots and generate $18$ order-$2$ symbols. The $36$ symbols can be recovered if the $18$ order-$2$ symbols are successfully delivered. The sum DoF can be expressed as
\begin{IEEEeqnarray}{rcl}
d_1^{mat}(2{,}3{,}2{,}3)&{=}&\frac{36}{6{+}18/d_2(3{,}2{,}3)}{.}\label{eq:d1_1_recur_323}
\end{IEEEeqnarray}

\subsubsection{Phase $1$, RT-PIN}\label{sec:rtpin323}
Here, before going into the scheme, let us briefly revisit the RT-PIN approach proposed in \cite{Abdoli13} for the $(1{,}1{,}3)$ IC. All the transmitters are co-scheduled for $5$ slots, during which each Tx sends $4$ symbols to the corresponding user. Such a transmission is termed as a redundancy transmission as the interference originated from a certain interferer spans a subspace of the received signal. Hence, by allevating the $4$ symbols of Rx2 (resp. Rx3), Rx1 is able to obtain a linear observation of its desired symbols only with interferer Rx3 (resp. Rx2). Then, the remaining overheard interferences in these two linear observations are considered as order-2 symbols. This process is known as RT-PIN.

In the $(3{,}2{,}3)$ IC, a trivial option is to switch off one antenna at each transmitter and perform a scaled version of the above scheme (like in a $(2{,}2{,}3)$ IC). However, such an option does not exploit the full benefit of the transmit antenna array. To counter this problem, we interpret the RT designed for the $(1{,}1{,}3)$ IC by two stages. The first stage, i.e., slot $1$ to $4$, is termed as the \emph{interference sensing} stage where each Tx identifies the row space spanned by each interference term. The second stage, i.e., the $5$th slot, is termed as the \emph{redundancy transmission} stage, where each Tx transmits a ``redundant'' linear combination of the symbols sent in the first stage. Following this idea, in the $(3{,}2{,}3)$ IC, we consider that
\begin{itemize}
\item in the first stage, there are $t_1{=}8$ slots and all the transmitters are co-scheulded, each of which transmits $3{\times}t_1{=}24$ symbols to the corresponding user;
\item in the second stage, there are $t_2{=}3$ slots and the symbols sent in the first stage are retransmitted. We note that in the $(3{,}2{,}3)$ IC, employing random precoders in the second stage does not yield a redundancy transmission. This is because each user has only $2$ antennas and the total number of linearly independent observation is $22$, which is smaller than the number of transmitted symbols per user. To solve this problem, the precoders in this stage are designed using perfect delayed CSIT to force each interference term into the $2{\times}t_1{=}16$ dimensional row space created in the first stage, so as to create a certain level of ``redundancy'' in the overheard interference;
\item by performing PIN, we obtain totally $36$ order-$2$ symbols to be delivered in phase $2$.
\end{itemize}
Specifically, the scheme operates as follows.

\underline{\emph{Interference sensing stage:}} According to the first bullet above, the received signal writes as
\begin{IEEEeqnarray}{rcl}
\mathbf{y}_k^{is}&{=}&\sum_{j{=}1{,}2{,}3}\underbrace{\text{\rm Bdiag}\left\{\mathbf{H}_{kj}(1){,}\cdots{,}\mathbf{H}_{kj}(8)\right\}\mathbf{W}_{j}^{is}}_{\mathbf{G}_{kj}^{is}}
\mathbf{u}_j{,}k{=}1{,}2{,}3,\label{eq:ykISK3}
\end{IEEEeqnarray}
where $\mathbf{u}_j{\in}\mathbb{C}^{24{\times}1}$, while $\mathbf{W}_{j}^{is}$ is a $24{\times}24$ full rank precoder across the $8$ time slots in this stage. The effective channel matrix is denoted by $\mathbf{G}_{kj}^{is}$ of size $16{\times}24$. Note that the superscript ``is'' stands for ``interference sensing''. From \eqref{eq:ykISK3}, we see that each interference term $\mathbf{G}_{kj}^{is}\mathbf{u}_j{,}k{\neq}j$, spans the full $16$ dimensions of the row space of the received signal because $\mathbf{G}_{kj}^{is}$ is full rank almost surely.

\underline{\emph{Redundancy transmission stage:}} For convenience, let us focus on the precoder design at Tx1 as the other two transmitters follow the same footsteps. As the $24$ symbols are retransmitted and there are $t_2{=}3$ slots in this stage, we design the precoder $\mathbf{W}_1^{rt}$ of size $Mt_2{\times}24{=}9{\times}24$ such that
\begin{IEEEeqnarray}{rcl}
\mathbf{W}_1^{rt}&{\subseteq}&\text{\rm rowspan}\left\{\mathbf{G}_{21}^{is}\right\}\cap\text{\rm rowspan}\left\{\mathbf{G}_{31}^{is}\right\}{,}\label{eq:WkK3case2rowspace}
\end{IEEEeqnarray}
where the superscript ``rt'' is short for ``redundancy transmission''. In this way, the linear space experienced by $\mathbf{u}_1$ at Rx$2$ (resp. Rx$3$) in the second stage, i.e., $\bar{\mathbf{H}}_{21}^{rt}\mathbf{W}_1^{rt}$ (resp. $\bar{\mathbf{H}}_{31}^{rt}\mathbf{W}_1^{rt}$), where $\bar{\mathbf{H}}_{21}^{rt}{\triangleq}\text{\rm Bdiag}\left\{\mathbf{H}_{21}(t)\right\}_{t{=}9{,}10{,}11}$ (resp. $\bar{\mathbf{H}}_{31}^{rt}{\triangleq}\text{\rm Bdiag}\left\{\mathbf{H}_{31}(t)\right\}_{t{=}9{,}10{,}11}$), will fall into the $16$-dimensional row space $\mathbf{G}_{21}^{is}$ (resp. $\mathbf{G}_{31}^{is}$).

Toward this, we firstly obtain a matrix $\mathbf{V}_1$ as $\mathbf{V}_1{=}\mathbf{D}_{21}\mathbf{G}_{21}^{is}{=}\mathbf{D}_{31}\mathbf{G}_{31}^{is}$, where $\mathbf{D}_{21}$ and $\mathbf{D}_{31}$ can be computed by
\begin{IEEEeqnarray}{rcl}
\left[\mathbf{D}_{21}\,\mathbf{D}_{31}\right]\left[\begin{array}{c}\mathbf{G}_{21}^{is}\\ -\mathbf{G}_{31}^{is}\end{array}\right]&{=}&\mathbf{0}.
\label{eq:K3Vk}
\end{IEEEeqnarray}
Since $\mathbf{G}_{21}^{is}$ and $\mathbf{G}_{31}^{is}$ are of size $16{\times}24$, their staggered matrix has a $8$-dimensional left null space almost surely. Then, both of $\mathbf{D}_{21}$ and $\mathbf{D}_{31}$ have size $8{\times}16$, thereby $\mathbf{V}_1$ has size $8{\times}24$ and is full rank, almost surely. Secondly, we compute $\mathbf{W}_1^{rt}$ as $\mathbf{W}_1^{rt}{=}\mathbf{C}_1\mathbf{V}_1$, where $\mathbf{C}_1$ of size $9{\times}8$ is a full rank mapping matrix.

\underline{\emph{PIN:}} Let us focus on Rx1 and write the received signal as
\begin{IEEEeqnarray}{rcl}
\!\!\!\!\!\!\mathbf{y}_1&{=}&\left[\!\!\begin{array}{ccc}
\mathbf{G}_{11}^{is}\!\! & \mathbf{G}_{12}^{is}\!\! & \mathbf{G}_{13}^{is}\\
\mathbf{H}_{11}^{rt}\mathbf{W}_1^{rt}\!\! & \mathbf{H}_{12}^{rt}\mathbf{W}_2^{rt}\!\! & \mathbf{H}_{13}^{rt}\mathbf{W}_3^{rt}\end{array}\!\!\right]\mathbf{u}{=}\left[\!\!\begin{array}{ccc}
\mathbf{G}_{11}^{is}\!\! & \mathbf{G}_{12}^{is}\!\! & \mathbf{G}_{13}^{is}\\
\mathbf{H}_{11}^{rt}\mathbf{W}_1^{rt}\!\! & \mathbf{H}_{12}^{rt}\mathbf{C}_2\mathbf{D}_{12}\mathbf{G}_{12}^{is}\!\! & \mathbf{H}_{13}^{rt}\mathbf{C}_3\mathbf{D}_{13}\mathbf{G}_{13}^{is}\end{array}\!\!\right]\mathbf{u}{,}\label{eq:y1forPINK3}
\end{IEEEeqnarray}
where $\mathbf{u}{\triangleq}{\rm stack}\{\mathbf{u}_1{,}\mathbf{u}_2{,}\mathbf{u}_3\}$. Then, we can see that the sub-matrix associated with each interference symbol vector has $N{\times}(t_1{+}t_2){=}22$ rows, but they only span the $16$-dimensional space of the first $16$ rows due to the redundancy transmission. Hence, there exists a $6{\times}22$ matrix $\mathbf{Q}_{1j}$, $j{=}2{,}3$, such that
\begin{IEEEeqnarray}{rcl}
\mathbf{Q}_{1j}\left[\begin{array}{c}\mathbf{G}_{1j}^{is}\\ \mathbf{H}_{1j}^{rt}\mathbf{C}_j\mathbf{D}_{1j}\mathbf{G}_{1j}^{is}\end{array}\right]&{=}&\mathbf{0},\label{eq:Q1jK3case2}
\end{IEEEeqnarray}
almost surely. Left-multiplying the received signal $\mathbf{y}_1$ by $\mathbf{Q}_{12}$ and $\mathbf{Q}_{13}$, we have
\begin{IEEEeqnarray}{rcl}
\bar{\mathbf{y}}_{1{,}2}&{=}&
\underbrace{\mathbf{Q}_{12}\left[\begin{array}{c}\mathbf{G}_{11}^{is}\\ \mathbf{H}_{11}^{rt}\mathbf{W}_1^{rt}\end{array}\right]}_{\bar{\mathbf{G}}_{11{,}2}{\in}\mathbb{C}^{6{\times}24}}\mathbf{u}_1{+}
\underbrace{\mathbf{Q}_{12}\left[\begin{array}{c}\mathbf{G}_{13}^{is}\\ \mathbf{H}_{13}^{rt}\mathbf{C}_3\mathbf{D}_{13}\mathbf{G}_{13}^{is}\end{array}\right]}_{ \bar{\mathbf{G}}_{13{,}2}{\in}\mathbb{C}^{6{\times}24}}\mathbf{u}_3{,}
\IEEEyessubnumber\label{eq:ybar12_rtpin_ph1}\\
\bar{\mathbf{y}}_{1{,}3}&{=}&
\underbrace{\mathbf{Q}_{13}\left[\begin{array}{c}\mathbf{G}_{11}^{is}\\ \mathbf{H}_{11}^{rt}\mathbf{W}_1^{rt}\end{array}\right]}_{\bar{\mathbf{G}}_{11{,}3}{\in}\mathbb{C}^{6{\times}24}}\mathbf{u}_1{+}
\underbrace{\mathbf{Q}_{13}\left[\begin{array}{c}\mathbf{G}_{12}^{is}\\ \mathbf{H}_{12}^{rt}\mathbf{C}_2\mathbf{D}_{12}\mathbf{G}_{12}^{is}\end{array}\right]}_{ \bar{\mathbf{G}}_{12{,}3}{\in}\mathbb{C}^{6{\times}24}}\mathbf{u}_2{,}
\IEEEyessubnumber\label{eq:ybar13_rtpin_ph1}
\end{IEEEeqnarray}
respectively. Then, Rx1 obtains $12$ observations, where $6$ observations are interfered with $\mathbf{u}_3$, while the remaining $6$ observations are interfered with $\mathbf{u}_2$. A similar approach is applied by Rx2 and Rx3.

\underline{\emph{Decoding feasibility:}} If $\bar{\mathbf{G}}_{13{,}2}\mathbf{u}_3$ in \eqref{eq:ybar12_rtpin_ph1} and $\bar{\mathbf{G}}_{12{,}3}\mathbf{u}_2$ in \eqref{eq:ybar13_rtpin_ph1} are removed and the side information obtained by Rx2 and Rx3, i.e., $\bar{\mathbf{G}}_{21{,}3}\mathbf{u}_1$ and $\bar{\mathbf{G}}_{31{,}2}\mathbf{u}_1$ respectively, are provided to Rx1, Rx1 has $24$ interference-free linear combinations of its $24$ desired symbols, i.e., ${\rm stack}\{\bar{\mathbf{G}}_{11{,}2}\,\bar{\mathbf{G}}_{11{,}3}\,\bar{\mathbf{G}}_{21{,}3}\,\bar{\mathbf{G}}_{31{,}2}\}\mathbf{u}_1$. These $24$ linear combinations are independent of each other. The proof is shown in Appendix C. Consequently, the $6$ elements in $\bar{\mathbf{G}}_{jk{,}l}\mathbf{u}_k{\in}\mathbb{C}^{6{\times}1}$ (which are made up of message of Rx$k$ and are obtained by Rx$j$ by nulling out the message of Rx$l$), for $k{\neq}j{\neq}l$, are order-$2$ symbols desired by Rx$k$ and Rx$j$, and thus denoted by $\mathbf{u}[k{|}k{,}j]{\in}\mathbb{C}^{6{\times}1}$.

To sum up, there are $72$ symbols transmitted in $11$ slots, generating $36$ order-$2$ symbols in total (each user has $12$ pieces of side information to be retransmitted). The sum DoF can be expressed as
\begin{IEEEeqnarray}{rcl}
d_1^{rt{-}pin}(3{,}3{,}2{,}3)&{=}&\frac{72}{11{+}36/d_2(3{,}2{,}3)}{.}\label{eq:d1_2_recur_323}
\end{IEEEeqnarray}
The remaining work is to calculate $d_2(3{,}2{,}3)$, which is discussed next.

\subsubsection{Phase $2$}
To propose a transmission strategy for phase $2$ of $(3{,}2{,}3)$ IC, let us briefly revisit the approach designed for the $(1{,}1{,}3)$ IC in \cite{Abdoli13}. To be specific, in the $(1{,}1{,}3)$ IC, Tx$1$ and Tx$2$ are co-scheduled for four slots. In the first two slots, Tx$1$ sends $2$ order-$2$ symbols to Rx1 and Rx2, while Tx$2$ sends one order-$2$ symbols to Rx1 and Rx2. Consequently, in the two-dimensional received signal at Rx3, the symbol sent by Tx$2$ spans only $1$ dimension, allowing Rx3 to perform PIN so as to obtain a linear observation purely of the symbols sent by Tx1. After that, in the third and fourth slot, we switch the role of Tx1 and Tx2, so that Rx3 obtains a linear observation purely of the symbols sent by Tx2. Those linear observations are useful for Rx1 and Rx2 and can be used to create future transmission in phase $3$. Next, following the same philosophy, we present how RT-PIN is performed in phase $2$ of the $(3{,}2{,}3)$ IC. We aim to transmit $42$ order-$2$ symbols in $12$ slots, which generate $9$ order-$3$ and $9$ order-$(1{,}2)$ symbols.

\underline{\emph{Redundancy Transmission:}} We consider that Tx$1$ and Tx$2$ are active for $4$ slots, where in the first two slots, Tx$1$ sends $6$ order-$2$ symbols, i.e., $\mathbf{u}[1{|}1{,}2]{\in}\mathbb{C}^{6{\times}1}$ while Tx$2$ sends $1$ order-$2$ symbol. The received signal at Rx$k{,}k{=}1{,}2{,}3$, writes as
\begin{IEEEeqnarray}{rcl}
\mathbf{y}_k{=}{\rm stack}\left\{\mathbf{y}_k(1){,}\mathbf{y}_k(2)\right\}&{=}&\mathbf{G}_{k1}\mathbf{u}[1|1{,}2]{+}\mathbf{G}_{k2}u[2|1{,}2]{,}
\label{eq:yk_323phase2}
\end{IEEEeqnarray}
where $\mathbf{G}_{k1}{\triangleq}\text{\rm Bdiag}\{\mathbf{H}_{k1}(1)\,\mathbf{H}_{k1}(2)\}\mathbf{W}_1$ of size $4{\times}6$ and $\mathbf{G}_{k2}{\triangleq}\text{\rm Bdiag}\{\mathbf{H}_{k2}(1)\,\mathbf{H}_{k2}(2)\}\mathbf{W}_2$ of size $4{\times}1$ follow the same notation as in \eqref{eq:y_323MATphase1}, while $\mathbf{W}_1$ and $\mathbf{W}_2$ are precoders of size $6{\times}6$ and $6{\times}1$, respectively. The time indexes refer to the first and second slot of phase 2.

At this moment, both Rx1 and Rx2 obtain $4$ linearly independent observations of the desired $7$ order-$2$ symbols transmitted by Tx1 and Tx2, thus requiring another $3$ linearly independent combinations to enable decoding. Toward this, according to \eqref{eq:yk_323phase2}, we see that Rx3 have $3$ redundant linear observations of $u[2|1{,}2]$. In other words, the dimension of the received signal, i.e., $4$, is greater than the size of $u[2|1{,}2]$, i.e., $1$, so that there exists a $3$ dimensional null space in $\mathbf{G}_{k2}$. This fact allows Rx3 to alleviate $u[2{|}1{,}2]$, thus obtaining $3$ linear combinations purely of $\mathbf{u}[1{|}1{,}2]$. In this way, the purified side information obtained by Rx3 after nulling out $\mathbf{u}[2{|}1{,}2]$ can be constructed by Tx1 and employed to formulate order-$3$ symbols.

\underline{\emph{PIN:}} Motivated by this, the PIN is conducted as,
\begin{IEEEeqnarray}{rcl}
\bar{\mathbf{y}}_{3{,}2}&{=}&\mathbf{F}_{32}\mathbf{y}_3{=}
\underbrace{\mathbf{F}_{32}\mathbf{G}_{31}\mathbf{u}[1{|}1{,}2]}_{\mathbf{u}[1{|}1{,}2{;}3]{\in}\mathbb{C}^{3{\times}1}},\label{eq:y3ph2case2}
\end{IEEEeqnarray}
where $\mathbf{F}_{32}{\in}\mathbb{C}^{3{\times}4}$ is such that $\mathbf{F}_{32}\mathbf{G}_{32}{=}\mathbf{0}$. Then, if the $3$-dimensional vector $\bar{\mathbf{y}}_{3{,}2}$ is provided to Rx1 and Rx2, both of Rx1 and Rx2 have $7$ linear combinations of the desired order-$2$ symbols as
\begin{IEEEeqnarray}{rcl}
\!\!\!\!\left[\!\!\begin{array}{c}\mathbf{y}_k\\ \bar{\mathbf{y}}_{3{,}2}\end{array}\!\!\right]&{=}&\left[\!\!\begin{array}{c}\mathbf{G}_{k1}\\ \mathbf{F}_{32}\mathbf{G}_{31}\end{array}\!\!\right]\mathbf{u}[1|1{,}2]{+}\!\!\left[\!\!\begin{array}{c}\mathbf{G}_{k2}\\ \mathbf{0}\end{array}\!\!\right]u[2|1{,}2],k{=}1{,}2{.}\label{eq:y12ph2case2}
\end{IEEEeqnarray}
The linear independence of these $7$ linear combinations are shown by the general proof in Appendix D.

The transmission in the third and fourth slot of phase 2 follows similarly by switching the role of Tx1 and Tx2. Specifically, Tx1 transmits $1$ order-$2$ symbols while Tx2 transmits $6$ order-$2$ symbols to Rx1 and Rx2. Then, after PIN, Rx3 obtains $3$ linear combinations purely of the order-$2$ symbols transmitted by Tx2, denoted by $\mathbf{u}[2|1{,}2{;}3]$. Moreover, in the four slots where Tx1 and Tx3 are active, Rx2 obtains $3$-dimensional vectors $\mathbf{u}[1|1{,}3{;}2]$ and $\mathbf{u}[3|1{,}3{;}2]$, while in the four slots where Tx2 and Tx3 are active, Rx1 obtains $3$-dimensional vectors $\mathbf{u}[2|2{,}3{;}1]$ and $\mathbf{u}[3|2{,}3{;}1]$. Thus, totally $42$ order-$2$ symbols are transmitted in $12$ slots, generating $18$ pieces of useful side information to be retransmitted.

The generation of the order-$3$ and order-$(1{,}2)$ symbols follow the footsteps designed for the $(3{,}1{,}3)$ IC. Recall that in Section \ref{sec:K3phase3}, $6$ pieces of side information result in $3$ order-$3$ and $3$ order-$(1{,}2)$ symbols. Now, since we have $18$ pieces of side information (scaled by the number of transmit antennas), we need $9$ order-$3$ and $9$ order-$(1{,}2)$ symbols. Then, the sum DoF of delivering order-$2$ symbols is expressed as
\begin{IEEEeqnarray}{rcl}
d_2(3{,}2{,}3)&{=}&\frac{42}{12{+}\frac{9}{d_3(M{,}N{,}3)}{+}\frac{9}{d_{1{,}2}(M{,}N{,}3)}}.
\label{eq:d2_323_recur}
\end{IEEEeqnarray}

\subsubsection{Phase $3$}
Since order-$3$ symbols are desired by all the three users and each user is equipped with $2$ antennas, the number of order-$3$ symbols that can be successfully transmitted and decoded per slot is $2$, thus $d_3(M{,}N{,}3){=}2$. After that, since the order-$(1{,}2)$ symbols for the three users can be transmitted simultaneously and each receiver is equipped with $2$ antennas, the number of order-$3$ symbols that can be successfully transmitted and decoded per slot is $6$, thus $d_{1{,}2}(M{,}N{,}3){=}6$. Substituting those quantities into \eqref{eq:d2_323_recur} yields $d_2(3{,}2{,}3){=}\frac{7}{3}$. Moreover, replacing $d_2(3{,}2{,}3){=}\frac{7}{3}$ into \eqref{eq:d1_1_recur_323} and \eqref{eq:d1_2_recur_323} leads to $d_1^{mat}(2{,}3{,}2{,}3){=}\frac{21}{8}$ and $d_1^{rt{-}pin}(3{,}3{,}2{,}3){=}\frac{504}{185}$, respectively.


%% file: RTPIN.tex
Here, with a $n$-transmitter/$n$-user scheduling, for $3{\leq}n{\leq}K$, we propose a general RT-PIN scheme in phase $1$ that achieves the sum DoF stated in Proposition \ref{theo:RTPIN}. Besides, we consider the case $\frac{M}{N}{\leq}\frac{1{+}(n{-}1)^2}{1{+}(n{-}2)(n{-}1)}$ as the achievability in the other case follows similarly by switching off the redundant transmit antennas. Without loss of generality, let us consider that the subset $\mathcal{S}_n{=}\{1{,}\cdots{,}n\}$ of uses are scheduled.

The RT-PIN is accomplished in two stages. Let us consider that the first stage, i.e., interference sensing stage, lasts for $t_1$ slots, while the second stage, i.e., redundancy transmission stage, lasts for $t_2$ slots. The values of $t_1$ and $t_2$ will be determined later on. The transmission strategy operates as follows.

\underline{\emph{Interference sensing stage:}} Each scheduled transmitter delivers $Mt_1$ symbols to the corresponding user. The received signal writes as
\begin{IEEEeqnarray}{rcl}
\mathbf{y}_k^{is}&{=}&\sum_{j{\in}\mathcal{S}_n}\underbrace{\text{\rm Bdiag}\left\{\mathbf{H}_{kj}(t)\right\}_{t{=}1{,}\cdots{,}t_1}{\times}\mathbf{W}_j^{is}}_{\mathbf{G}_{kj}^{is}{\in}\mathbb{C}^{Nt_1{\times}Mt_1}}
\mathbf{u}_j{,}k{\in}\mathcal{S}_n,\label{eq:ykIS}
\end{IEEEeqnarray}
where $\mathbf{u}_j{\in}\mathbb{C}^{Mt_1{\times}1}$ and $\mathbf{W}_j^{is}$ is a $Mt_1{\times}Mt_1$ full rank precoder. From \eqref{eq:ykIS}, we see that each interference term $\mathbf{G}_{kj}^{is}\mathbf{u}_j{,}k{\neq}j$ spans the full $Nt_1$ dimension of the row space of the received signal because $N{\leq}M$ and $\mathbf{G}_{kj}^{is}$ is full rank almost surely.

\underline{\emph{Redundancy transmission stage:}} The objective in the second stage is to design precoders that force each overheard interference into the $Nt_1$-dimensional row space created in the first stage. Without loss of generality, let us consider the precoder design at Tx1. We aim to design $\mathbf{W}_1^{rt}{\in}\mathbb{C}^{Mt_2{\times}Mt_1}$ such that
\begin{IEEEeqnarray}{rcl}
\mathbf{W}_1^{rt}&{\subseteq}&\text{\rm rowspan}\left\{\mathbf{G}_{21}^{is}\right\}\cap\text{\rm rowspan}\left\{\mathbf{G}_{31}^{is}\right\}\cap{\cdots}\cap\text{\rm rowspan}\left\{\mathbf{G}_{n1}^{is}\right\}{.}\label{eq:Wkcase2rowspace}
\end{IEEEeqnarray}
Following the footstep in Section \ref{sec:rtpin323}, we firstly obtain a matrix $\mathbf{V}_1$ such that $\mathbf{V}_1{=}\mathbf{D}_{21}\mathbf{G}_{21}^{is}{=}{\cdots}{=}\mathbf{D}_{n1}\mathbf{G}_{n1}^{is}$, where $\mathbf{D}_{j1}{,}j{=}2{,}\cdots{,}n$, is computed by,
\begin{IEEEeqnarray}{rcl}
\left[\mathbf{D}_{21}\,\mathbf{D}_{31}\,\cdots\,\mathbf{D}_{n1}\right]
\underbrace{\left[\begin{array}{cccc}\mathbf{G}_{21}^{is} & \mathbf{0} & \cdots & \mathbf{0}\\
-\mathbf{G}_{31}^{is} & \mathbf{G}_{31}^{is} & \ddots & \mathbf{0}\\
\mathbf{0} & -\mathbf{G}_{41}^{is} & \ddots & \mathbf{0}\\
\vdots & \ddots & \ddots & \mathbf{G}_{(n{-}1)1}^{is}\\
\mathbf{0} & \cdots & \cdots & -\mathbf{G}_{n1}^{is}\end{array}\right]}_{\boldsymbol\Phi_1{\in}\mathbb{C}^{(n{-}1)N{\times}(n{-}1)M}}&{=}&\mathbf{0}.
\label{eq:DG}
\end{IEEEeqnarray}
Due to the fact $\frac{M}{N}{\leq}\frac{1{+}(n{-}1)^2}{1{+}(n{-}2)(n{-}1)}{<}\frac{n{-}1}{n{-}2}$, the dimension of the left null space of $\boldsymbol\Phi_1$ is $[(n{-}1)N{-}(n{-}2)M]t_1$. Then, $\mathbf{D}_{j1}$, for $j{=}2{,}\cdots{,}n$, has size $[(n{-}1)N{-}(n{-}2)M]t_1{\times}Nt_1$ and $\mathbf{V}_1$ is a $[(n{-}1)N{-}(n{-}2)M]t_1{\times}Mt_1$ matrix and is full rank, i.e., $[(n{-}1)N{-}(n{-}2)M]t_1$, almost surely. Secondly, we obtain the precoding matrix as $\mathbf{W}_1^{rt}{=}\mathbf{C}_1\mathbf{V}_1$, where $\mathbf{C}_1{\in}\mathbb{C}^{Mt_2{\times}((n{-}1)N{-}(n{-}2)M)t_1}$ is a full rank mapping matrix.

\underline{\emph{PIN:}} Thanks to the redundancy transmission, each user is able to perform PIN. Let us focus on Rx$1$ for convenience. The received signal writes as
\begin{IEEEeqnarray}{rcl}
\!\!\!\!\!\!\mathbf{y}_1&{=}&\left[\!\!\begin{array}{cccc}
\mathbf{G}_{11}^{is}\!\! & \mathbf{G}_{12}^{is} & \cdots\!\! & \mathbf{G}_{1n}^{is}\\
\mathbf{H}_{11}^{rt}\mathbf{W}_1^{rt}\!\! & \mathbf{H}_{12}^{rt}\mathbf{C}_2\mathbf{D}_{12}\mathbf{G}_{12}^{is} & \cdots & \mathbf{H}_{1n}^{rt}\mathbf{C}_n\mathbf{D}_{1n}\mathbf{G}_{1n}^{is}\end{array}\!\!\right]\!\!\mathbf{u}{,}
\end{IEEEeqnarray}
where $\mathbf{u}{\triangleq}{\rm stack}\left\{\mathbf{u}_1{,}\cdots\mathbf{u}_n\right\}$. Clearly, there exists a $Nt_2{\times}N(t_1{+}t_2)$ matrix $\mathbf{Q}_{1j}{,}j{=}2{,}\cdots{,}n$, such that
\begin{IEEEeqnarray}{rcl}
\mathbf{Q}_{1j}\left[\begin{array}{c}\mathbf{G}_{1j}^{is}\\ \mathbf{H}_{1j}^{rt}\mathbf{C}_{j}\mathbf{D}_{1j}\mathbf{G}_{1j}^{is}\end{array}\right]&{=}&\mathbf{0}.\label{eq:Qi1ij}
\end{IEEEeqnarray}
Then, left-multiplying $\mathbf{y}_1$ by $\mathbf{Q}_{1j}$ yields
\begin{IEEEeqnarray}{rcl}
\bar{\mathbf{y}}_{1{,}j}&{=}&\underbrace{\mathbf{Q}_{1j}\left[\begin{array}{c}\mathbf{G}_{11}^{is}\\ \mathbf{H}_{11}^{rt}\mathbf{W}_1^{rt}\end{array}\right]}_{\bar{\mathbf{G}}_{11{,}j}{\in}\mathbb{C}^{Nt_2{\times}Mt_1}}\mathbf{u}_1{+}
\sum_{l{\in}\mathcal{S}_n{\setminus}\{1{,}j\}}\underbrace{\mathbf{Q}_{1j}\left[\begin{array}{c}\mathbf{G}_{1l}^{is}\\ \mathbf{H}_{1l}^{rt}\mathbf{W}_l^{rt}\end{array}\right]}_{\bar{\mathbf{G}}_{1l{,}j}{\in}\mathbb{C}^{Nt_2{\times}Mt_1}}\mathbf{u}_l.
\label{eq:yeffectcase2ph1}
\end{IEEEeqnarray}

In $\bar{\mathbf{y}}_{1{,}j}$, Rx$1$ obtains $Nt_2$ observations of the desired symbols mixed with $n{-}2$ interferers, i.e., symbols of Rx$l$, $l{\in}\mathcal{S}_n{\setminus}\{1{,}j\}$. Since there are $n{-}1$ choices of $\mathbf{Q}_{1j}$, the total number of interferences overheard (after PIN) by Rx$1$ is $(n{-}1)(n{-}2)Nt_2$, while there are $(n{-}1)Nt_2$ observations of the desired symbols.

\underline{\emph{Decoding feasibility:}} A similar PIN is performed by the other users. The interferences caused by $\mathbf{u}_1$ at Rx$l$ by nulling out the message of Rx$j$ are denoted by $\bar{\mathbf{G}}_{l1{,}j}\mathbf{u}_1{\in}\mathbb{C}^{Nt_2{\times}1}$, ${\forall}l{\neq}j{,}\{l{,}j\}{=}2{,}\cdots{,}n{-}1$. Since there are $(n{-}1)(n{-}2)$ possible choices of such $l$ and $j$, there are totally $(n{-}1)(n{-}2)Nt_2$ interference symbols made by $\mathbf{u}_1$. If all those pieces of side information are provided to Rx$1$ and all $\bar{\mathbf{G}}_{1l{,}j}\mathbf{u}_l$ in \eqref{eq:yeffectcase2ph1} are removed, Rx$1$ has $(n{-}1)^2Nt_2$ interference-free linear observations of the desired symbols, i.e.,
\begin{IEEEeqnarray}{rcl}
\!\!\!\!{\rm stack}\left\{\{\bar{\mathbf{G}}_{11{,}j}\}_{{\forall}j{=}2{,}\cdots{,}n{-}1}\,\,
\{\bar{\mathbf{G}}_{l1{,}j}\}_{{\forall}l{\neq}j{,}\{l{,}j\}{=}2{,}\cdots{,}n{-}1}\right\}\!\!\mathbf{u}_1.\label{eq:u1K}
\end{IEEEeqnarray}
It is shown in Appendix C that the above effective channel matrix is full rank almost surely, if
\begin{IEEEeqnarray}{rcl}
(n{-}1)^2Nt_2{=}Mt_1{,}&\quad\text{\rm and}\quad&
Mt_1{\leq}\left((n{-}1)^2{+}1\right)\!\left[(n{-}1)N{-}(n{-}2)M\right]t_1{.}\label{eq:constraints}
\end{IEEEeqnarray}
Since we consider $\frac{M}{N}{\leq}\frac{(n{-}1)^2{+}1}{(n{-}2)(n{-}1){+}1}$, we have $t_1{=}N(n{-}1)^2$ and $t_2{=}M$. When $\frac{M}{N}{>}\frac{(n{-}1)^2{+}1}{(n{-}2)(n{-}1){+}1}$, the above scheme is still feasible by using only $\hat{M}{\triangleq}\min\left\{M{,}\frac{1{+}(n{-}1)^2}{1{+}(n{-}2)(n{-}1)}N\right\}$ antennas at each Tx\footnote{If $\hat{M}$ is not an integer, we perform the scheme with a time-extension. For instance, when $K{=}n{=}3$ and $\hat{M}{=}\frac{5}{3}N$, we choose $t_2{=}3{\times}\hat{M}{=}5N$, $t_1{=}3{\times}4N{=}12N$. Each Tx sends $\hat{M}{\times}t_1{=}20N^2$ symbols and $6Nt_2{=}30N$ order-$2$ symbols are generated.} and choosing $t_1{=}N(n{-}1)^2$ and $t_2{=}\hat{M}$.

To sum up, since the transmission strategy is performed ${{K}\choose{n}}$ times for all the possible subsets of $n$ transmitters, the total number of symbols and slots are given by
\begin{IEEEeqnarray}{rcl}
N_1{=}nMt_1{{K}\choose{n}}{=}n(n{-}1)^2MN{{K}\choose{n}}{,}&\quad&
T_1{=}(t_1{+}t_2){{K}\choose{n}}{=}(N(n{-}1)^2{+}M){{K}\choose{n}}.\label{eq:N1T1_RTPIN}
\end{IEEEeqnarray}
Besides, considering the $(n{-}1)(n{-}2)Nt_2$ overheard interferences (see \eqref{eq:yeffectcase2ph1}) at each user as order-$2$ symbols, the total number of order-$2$ symbols generated in phase $1$ is
\begin{IEEEeqnarray}{rcl}
N_2&{=}&n(n{-}1)(n{-}2)Nt_2{{K}\choose{n}}{=}n(n{-}1)(n{-}2)MN{{K}\choose{n}}.\label{eq:N2_RTPIN}
\end{IEEEeqnarray}
As the achievable sum DoF can be expressed as $d_1^{rt{-}pin}(n{,}M{,}N{,}K){=}\frac{N_1}{T_1{+}N_2/d_2(M{,}N{,}K)}$, Theorem \ref{theo:RTPIN} is immediate with the parameters in \eqref{eq:N1T1_RTPIN} and \eqref{eq:N2_RTPIN}.

%% file: phasemMNK.tex
In this subsection, we aim to propose the transmission strategy to deliver order-$m$ ($2{\leq}m{\leq}K{-}1$) symbols in the $(M{,}N{,}K)$ IC with $1{\leq}\frac{M}{N}{\leq}K$. Following the $K$-phase framework illustrated in Figure \ref{fig:flow}, we can obtain the recursive expression of sum DoF of delivering order-$m$ symbols as
\begin{IEEEeqnarray}{rcl}
d_m(M{,}N{,}K)&{=}&\frac{N_m}{T_m{+}\frac{N_{m{+}1}}{d_{m{+}1}(M{,}N{,}K)}{+}\frac{N_{1{,}m}}{d_{1{,}m}(M{,}N{,}K)}}{,}2{\leq}m{\leq}K{-}1{,}
\label{eq:dofm_MNK_recur}\\
d_K(M{,}N{,}K)&{=}&N.\label{eq:dofK_MNK}
\end{IEEEeqnarray}
where $d_{1{,}m}(M{,}N{,}K)$ refers to the DoF of delivering order-$(1{,}m)$ symbols in the $(M{,}N{,}K)$ IC. Following the discussion in Section \ref{sec:mp1K1K}, since the order-$(1{,}m)$ symbols for a subset of $m{+}1$ users can be transmitted simultaneously and each receiver is equipped with $N$ antennas, we have $d_{1{,}m}(M{,}N{,}K){=}N(m{+}1)$. Besides, \eqref{eq:dofK_MNK} is due to the fact that order-$K$ symbols are intended for all users and each user is equipped with $N$ antennas. Hence, the work is reduced to propose transmission strategy for phase $m$-I, for $2{\leq}m{\leq}K{-}1$, and identify the parameters $N_m$, and $N_{m{+}1}$, $T_m$ and $N_{1{,}m}$ in \eqref{eq:dofm_MNK_recur}.

Next, we will focus on the case $M{\leq}N(K{-}m{+}1)$ because the scheme in the case $M{>}N(K{-}m{+}1)$ follows similarly by switching off the redundant transmit antennas.

Similar to the $(3{,}2{,}3)$ IC, we consider that the transmission duration is divided into ${{K}\choose{m}}$ rounds, each of which is dedicated to deliver order-$m$ symbols to a certain subset $\mathcal{S}_m$ of users. We sort the $m$ elements of $\mathcal{S}_m$ in an cyclic ascending order as $\mathcal{S}_m{\triangleq}\{i_1{,}\cdots{,}i_m\}$, where $i_1{<}\cdots{<}i_m$. Then, the $m$ transmitters are scheduled in a pair-wise manner\footnote{In the $(3{,}2{,}3)$ IC example presented in Section \ref{sec:ach323}, in round 1, the scheduling is Rx $(1{,}2)$ and Rx $(2{,}1)$.}, i.e., $(i_1{,}i_2)$, $(i_2{,}i_3)$, $\cdots$, $(i_m{,}i_1)$. This procedure is known as a $2$-transmitter/$m$-user scheduling. Next, without loss of generality, we focus on $\mathcal{S}_m{=}\{1{,}\cdots{,}m\}$ and consider that Tx$1$ and Tx$2$ are active.

\underline{\emph{Redundancy transmission:}} Tx$1$ and Tx$2$ are scheduled for $t{=}K{-}m{+}1$ slots, where Tx$1$ transmits $Mt$ order-$2$ symbols, i.e., $\mathbf{u}[1|\mathcal{S}_m]{\in}\mathbb{C}^{Mt{\times}1}$, and Tx$2$ transmits $Nt{-}M$ order-$2$ symbols, i.e., $\mathbf{u}[2|\mathcal{S}_m]{\in}\mathbb{C}^{(Nt{-}M){\times}1}$. The signal received by Rx$k$, for $k{=}1{,}\cdots{,}K$, writes as
\begin{IEEEeqnarray}{rcl}
\mathbf{y}_k&{=}&[\underbrace{\bar{\mathbf{H}}_{k1}\mathbf{W}_1}_{\mathbf{G}_{k1}{\in}\mathbb{C}^{Nt{\times}Mt}}{,}
\underbrace{\bar{\mathbf{H}}_{k2}\mathbf{W}_2}_{\mathbf{G}_{k2}{\in}\mathbb{C}^{Nt{\times}(Nt{-}M)}}]
{\rm stack}\left\{\mathbf{u}[1|\mathcal{S}_m]{,} \mathbf{u}[2|\mathcal{S}_m]\right\}{,}\label{eq:yk_phasemMNK}
\end{IEEEeqnarray}
where $\mathbf{W}_1$ and $\mathbf{W}_2$ are full rank precoding matrices with size $Mt{\times}Mt$ and $Mt{\times}(Nt{-}M)$, respectively, while $\bar{\mathbf{H}}_{kj}{=}\text{\rm Bdiag}\left\{\mathbf{H}_{kj}(1){,}\cdots{,}\mathbf{H}_{kj}(t)\right\}{,}j{=}1{,}2$, is the $Nt{\times}Mt$ channel matrix across the time slots.

At this moment, each scheduled user obtains $Nt$ linearly independent observations of the desired $M(t{-}1){+}Nt$ order-$2$ symbols, requiring another $M(t{-}1)$ linearly independent observations to enable the decoding. Toward this, we notice that for any user ${\forall}k{=}1{,}{\cdots}{,}K$, there exists a $M$-dimensional left null space of $\mathbf{G}_{ki_2}$ as $\mathbf{G}_{ki_2}$ is a $Nt{\times}(Nt{-}M)$ full rank matrix almost surely. This allows each non-scheduled user to null out $\mathbf{u}[2|\mathcal{S}_m]$ and attain $M$ linear observations purely of $\mathbf{u}[1|\mathcal{S}_m]$.

\underline{\emph{PIN:}} Specifically, for Rx$j{,}j{=}m{+}1{,}\cdots{,}K$, we have
\begin{IEEEeqnarray}{rcl}
\bar{\mathbf{y}}_{j{,}2}&{=}&\mathbf{F}_{j2}\mathbf{y}_j{=}
\underbrace{\mathbf{F}_{j2}\mathbf{G}_{j1}\mathbf{u}[1|\mathcal{S}_m]}_{\mathbf{u}[1|\mathcal{S}_m;j]{\in}\mathbb{C}^{M{\times}1}}{,}
j{=}m{+}1{,}\cdots{,}K.
\label{eq:PINphasemI}
\end{IEEEeqnarray}
where $\mathbf{F}_{j2}{\in}\mathbb{C}^{M{\times}Nt}$ is such that $\mathbf{F}_{j2}\mathbf{G}_{j2}{=}\mathbf{0}$. Then, we can see that if all the $M$-dimensional vectors $\bar{\mathbf{y}}_{j2}$, $j{=}m{+}1{,}\cdots{,}K$, are provided to the scheduled users, each scheduled user obtains totally $Nt{+}M(K{-}m){=}Nt{+}M(t{-}1)$ linear observations as
\begin{IEEEeqnarray}{rcl}
\!\!\!\!\!\!\left[\!\!\begin{array}{c}\mathbf{y}_k\\ \bar{\mathbf{y}}_{m{+}1{,}2}\\ \vdots\\ \bar{\mathbf{y}}_{K{,}2}\end{array}\!\!\right]&{=}&\left[\!\!\begin{array}{cc}\mathbf{G}_{k1} & \!\!\!\!\!\!\mathbf{G}_{k2}\\ \mathbf{F}_{(m{+}1)2}\mathbf{G}_{(m{+}1)1} & \!\!\!\!\!\!\mathbf{0}\\ \vdots & \!\!\!\!\!\!\vdots\\ \mathbf{F}_{K2}\mathbf{G}_{K1} & \!\!\!\!\!\!\mathbf{0} \end{array}\!\!\!\!\right]\!\!\left[\!\!\begin{array}{c}\mathbf{u}[1{|}\mathcal{S}_m]\\ \mathbf{u}[2{|}\mathcal{S}_m]\end{array}\!\!\right]\!\!{,}k{=}1{,}\cdots{,}m{.}\label{eq:yKph2_effect}
\end{IEEEeqnarray}
These $Nt{+}M(t{-}1)$ observations are shown to be linearly independent in Appendix D. In this way, the purified signal at the non-scheduled user, i.e., $\mathbf{u}[1|\mathcal{S}_m;j]$, is able to be reconstructed by Tx$1$ at the end of phase $m$-I, and can be considered as useful signals for the scheduled users.

Similarly, when Tx$2$ and Tx$3$ are scheduled, another $Nt{+}M(t{-}1)$ order-$m$ symbols are transmitted and $M(K{-}m)$ useful signals $\mathbf{u}[2|\mathcal{S}_m;j]$, for $j{=}m{+}1{,}\cdots{,}K$, are generated. Then, as the same transmissions strategy is employed $m$ times per round and there are ${{K}\choose{m}}$ rounds in phase $m$-I, we have
\begin{IEEEeqnarray}{rcl}
N_m{=}\left[Nt{+}M(t{-}1)\right]m{{K}\choose{m}}{,}&\quad&
T_m{=}tm{{K}\choose{m}},\label{eq:NmTm}
\end{IEEEeqnarray}
where $t{=}K{-}m{+}1$. Besides, the total number of resultant useful signals is $mM(K{-}m){{K}\choose{m}}$. Next, we employ these useful signals to formulate order-$(m{+}1)$ and order-$(1{,}m)$ symbols.

The generation of the order-$(m{+}1)$ and order-$(1{,}m)$ symbols follow the footsteps designed for the $(K{,}1{,}K)$ IC. The only difference lies in that each term shown in Table \ref{tab:ordermp1} becomes a $M{\times}1$ vector in the $(M{,}N{,}K)$ IC. Hence, the number of order-$(m{+}1)$ and order-$(1{,}m)$ symbols should be scaled by $M$. Specifically, we have
\begin{IEEEeqnarray}{rcl}
N_{m{+}1}{=}M(m{-}1)(m{+}1){{K}\choose{m{+}1}},&\quad&
N_{1{,}m}{=}M(m{+}1){{K}\choose{m{+}1}}.\label{eq:Nmp1N1m}
\end{IEEEeqnarray}

Then, plugging \eqref{eq:NmTm} and \eqref{eq:Nmp1N1m} into \eqref{eq:dofm_MNK_recur}, the recursive expression of the DoF of delivering order-$m$ symbols, i.e., $d_m(M{,}N{,}K)$, are given by
\begin{IEEEeqnarray}{rcl}
d_m(M{,}N{,}K)&{=}&\frac{\left[(M^\prime{+}N)(K{-}m){+}N\right]m{{K}\choose{m}}}
{(K{-}m{+}1)m{{K}\choose{m}}{+}\frac{M^\prime(m{-}1)(m{+}1){{K}\choose{m{+}1}}}{d_{m{+}1}(M{,}N{,}K)}{+}
\frac{M^\prime}{N}{{K}\choose{m{+}1}}}.\label{eq:dofm_v2}
\end{IEEEeqnarray}
where $M^\prime{\triangleq}\min\{M{,}N(K{-}m{+}1)\}$. Theorem \ref{theo:orderm} holds following the derivations in Appendix B.
\begin{myremark}
\emph{Notably, we point out that when $M{\geq}(K{-}m{+}1)N{=}Nt$, the proposed scheme in phase $m$-I smoothly connects with the MAT-like transmission designed for the $(K{,}1{,}K)$ IC. Recall that in \eqref{eq:yk_phasemMNK}, the size of $\mathbf{u}[2{|}\mathcal{S}_m]$ becomes zero when $M{=}N(K{-}m{+}1){=}Nt$, implying that Tx$2$ becomes silent and only Tx$1$ is scheduled. Then, any matrix can be considered as lying in the null space of $\mathbf{G}_{k2}$. Consequently, there is no need to perform PIN and the overheard interferences $\mathbf{G}_{j1}\mathbf{u}[1{|}\mathcal{S}_m]$ can be regarded as useful signals. More specifically, the recursive equation \eqref{eq:dofm_v2} becomes \eqref{eq:dofm_K1K_recur2} if we replace $N{=}1$ and $M{=}K{-}m{+}1$.}
\end{myremark} 

%% file: AppICdelay_1col.tex
\small
\subsection{Derivation of the bounded value $\frac{64}{15}$ for $d_1(K{,}1{,}K)$ in Theorem \ref{theo:K1K}}
For convenience, we approximate $2\mathcal{O}(K){\approx}\mathcal{O}_{i^*}(K)$, where $i^*$ is the solution to \eqref{eq:d1_K1K}. For $K{\to}\infty$, to show $d_1(K{,}1{,}K){\approx}\frac{64}{15}$, it suffices to prove $\mathcal{O}(K){\approx}4$, namely $A_2(K{,}1{,}K){\approx}\frac{3}{4}$. From \eqref{eq:Am2}, we have
\begin{IEEEeqnarray}{rcl}
A_2(K{,}1{,}K)&{=}& \frac{K}{(K{-}1)}\sum_{l{=}2}^{K{-}1}\frac{1}{(l{-}1)(l{+}1)}{-}\frac{1}{K{-}1}\sum_{l{=}2}^{K{-}1}\frac{l}{(l{-}1)(l{+}1)}\nonumber\\
&\stackrel{K{\to}{\infty}}{\approx}&\frac{K}{2(K{-}1)}\sum_{l{=}2}^{K{-}1}\frac{1}{l{-}1}{-}\frac{1}{l{+}1}\nonumber\\ &{=}&\frac{K}{2(K{-}1)}(1{+}\frac{1}{2}{-}\frac{1}{K{-}1}{-}\frac{1}{K})\stackrel{K{\to}{\infty}}{\approx}\frac{3}{4}.
\end{IEEEeqnarray}

\subsection{Derivation of $d_m(M{,}N{,}K)$ using the recursive equation \eqref{eq:dofm_v2}}
The proof is shown considering the case 1) $M{\geq}(K{-}m{+}1)N$ and the case 2) $M{\leq}(K{-}m{+}1)N$.

When $M{\geq}(K{-}m{+}1)N$, we can rewrite \eqref{eq:dofm_v2} as $\frac{d_m(M{,}N{,}K)}{N}{=}\frac{m(K{-}m{+}1)}{m{+}\frac{K{-}m}{m{+}1}{+}\frac{(m{-}1)(K{-}m)}{\frac{d_m(M{,}N{,}K)}{N}}}$. Let us introduce $A_m^\prime{\triangleq}\\1{-}\frac{1}{\frac{d_m(M{,}N{,}K)}{N}}$ and $A_m$ in \eqref{eq:Am2} is obtained by $1{-}\frac{1}{N}{+}\frac{1}{N}A_m^\prime$. Then, for $K{-}\lfloor\frac{M}{N}\rfloor{+}1{\leq}m{\leq}K{-}1$, one has
\begin{equation}
A_m^\prime{=}\underbrace{\frac{(K{-}m)(m{-}1)}{m(K{-}m{+}1)}}_{B_m}A_{m{+}1}^\prime{+}
\underbrace{\frac{K{-}m}{(K{-}m{+}1)(m{+}1)}}_{C_m},\label{eq:Amp}
\end{equation}
followed by $B_mA_{m{+}1}^\prime{=}B_mB_{m{+}1}A_{m{+}2}^\prime{+}B_mC_{m{+}1}$, $B_mB_{m{+}1}A_{m{+}2}^\prime{=}B_mB_{m{+}1}B_{m{+}2}A_{m{+}3}^\prime{+}B_mB_{m{+}1}C_{m{+}2}$, till $A_{K{-}1}^\prime\prod_{i{=}m}^{K{-}2}B_i{=}A_K^\prime\prod_{i{=}m}^{K{-}1}B_i{+}C_{K{-}1}\prod_{i{=}m}^{K{-}2}B_i$, resulting in
\begin{equation}
A_m^\prime{=}A_K^\prime\prod_{i{=}m}^{K{-}1}B_i{+}\sum_{l{=}m}^{K{-}1}C_l\prod_{i{=}m}^{l{-}1}B_i.\label{eq:Amp2}
\end{equation}
By the definition of $B_i$ and $C_i$ \eqref{eq:Amp}, it is easily verified that
\begin{IEEEeqnarray}{rcl}
\prod_{i{=}m}^{K{-}1}B_i{=}\frac{m{-}1}{(K{-}1)(K{-}m{+}1)}&,&
C_l\prod_{i{=}m}^{l{-}1}B_i{=}\frac{m{-}1}{K{-}m{+}1}\frac{K{-}l}{l{+}1(l{-}1)}{.}\label{eq:sumBC}
\end{IEEEeqnarray}
Substituting \eqref{eq:sumBC} into \eqref{eq:Amp2} leads to \eqref{eq:Am2}.

When $M{\leq}(K{-}m{+}1)N$, by introducing $A_m{=}1{-}\frac{1}{d_m(M{,}N{,}K)}$, we have a recursive equation in terms of $A_m$ as
\begin{IEEEeqnarray}{rcl}
A_m&{=}&\underbrace{\frac{M(m{-}1)(K{-}m)}{m\left[(M{+}N)(K{-}m){+}N\right]}}_{B_m}A_{m{+}1}{+}
\underbrace{\frac{M(K{-}m){+}m(N{-}1)(K{-}m{+}1){-}\frac{(K{-}m)M}{(m{+}1)N}}{m\left[(M{+}N)(K{-}m){+}N\right]}}_{C_m}{.}\label{eq:Amrecur}
\end{IEEEeqnarray}
Then, following the footsteps of deriving \eqref{eq:Amp2}, we have
\begin{IEEEeqnarray}{rcl}
A_m&{=}&A_{K{-}\lfloor\frac{M}{N}\rfloor{+}1}\prod_{i{=}m}^{K{-}\lfloor\frac{M}{N}\rfloor}B_i{+}
\sum_{l{=}m}^{K{-}\lfloor\frac{M}{N}\rfloor}C_l\prod_{i{=}m}^{l{-}1}B_i{.}
\end{IEEEeqnarray}
By the definition of $B_m$ and $C_m$ in \eqref{eq:Amrecur}, it can be shown that $\Theta_m$ in \eqref{eq:Thetam} is obtained by $\prod_{i{=}m}^{K{-}\lfloor\frac{M}{N}\rfloor}B_i$, while $\Delta_m$ in \eqref{eq:Deltaml} is obtained by $C_l\prod_{i{=}m}^{l{-}1}B_i$. Besides, $A_{K{-}\lfloor\frac{M}{N}\rfloor{+}1}$ is obtained by replacing $m{=}K{-}\lfloor\frac{M}{N}\rfloor{+}1$ into \eqref{eq:Am2}. This completes the proof.

\subsection{Proof of the linear independence of the linear observations in \eqref{eq:u1K}}
According to \eqref{eq:yeffectcase2ph1}, let us write the submatrices in \eqref{eq:u1K} as
\begin{IEEEeqnarray}{rcl}
\bar{\mathbf{G}}_{11{,}j}{=}\mathbf{Q}_{1j}\left[\begin{array}{c}\mathbf{G}_{11}^{is} \\ \mathbf{H}_{11}^{rt}\mathbf{C}_1\mathbf{D}_{j1}\mathbf{G}_{11}^{is}\end{array}\right]{,} {\forall}j{=}2{,}\cdots{,}n{,}&\quad &
\bar{\mathbf{G}}_{l1{,}j}{=}\mathbf{Q}_{lj}\left[\begin{array}{c}\mathbf{G}_{l1}^{is} \\ \mathbf{H}_{l1}^{rt}\mathbf{C}_1\mathbf{D}_{l1}\mathbf{G}_{l1}^{is}\end{array}\right]{,} {\forall}\{l{,}j\}{=}2{,}\cdots{,}n{,}l{\neq}j{.}\label{eq:Gbi1i1ij_Gbili1ij}
\end{IEEEeqnarray}
Besides, due to \eqref{eq:Qi1ij}, $\mathbf{Q}_{lj}{,}{\forall}j{=}2{,}\cdots{,}n{,}{\forall}l{\in}1{,}\cdots{,}n{,}l{\neq}j$ can be expressed as $\mathbf{Q}_{lj}{=}\left[\mathbf{H}_{lj}^{rt}\mathbf{C}_j\mathbf{D}_{lj}\quad -\mathbf{I}\right]$
where $\mathbf{I}$ stands for identity matrix. Then, \eqref{eq:Gbi1i1ij_Gbili1ij} rewrites as
\begin{IEEEeqnarray}{rcl}
\bar{\mathbf{G}}_{11{,}j}&{=}&\mathbf{H}_{1j}^{rt}\mathbf{C}_j\mathbf{D}_{1j}\mathbf{G}_{11}^{is}{-} \mathbf{H}_{11}^{rt}\mathbf{C}_1\mathbf{D}_{j1}\mathbf{G}_{j1}^{is}{=}
\left[\mathbf{H}_{1j}^{rt}\mathbf{C}_j\,\,\,{-}\mathbf{H}_{11}^{rt}\mathbf{C}_1\right] \left[\begin{array}{c}\mathbf{D}_{1j}\mathbf{G}_{11}^{is} \\ \mathbf{D}_{j1}\mathbf{G}_{j1}^{is}\end{array}\right] {,}\IEEEyessubnumber\label{eq:Gbi1i1ij2}\\
\bar{\mathbf{G}}_{l1{,}j}&{=}&\mathbf{H}_{lj}^{rt}\mathbf{C}_j\mathbf{D}_{lj}\mathbf{G}_{l1}^{is}{-} \mathbf{H}_{l1}^{rt}\mathbf{C}_1\mathbf{D}_{l1}\mathbf{G}_{l1}^{is}
{=}\left[\mathbf{H}_{lj}^{rt}\mathbf{C}_j\,\,\,{-}\mathbf{H}_{l1}^{rt}\mathbf{C}_1\right] \left[\begin{array}{c}\mathbf{D}_{lj}\mathbf{G}_{l1}^{is} \\ \mathbf{D}_{l1}\mathbf{G}_{l1}^{is}\end{array}\right] {.}\IEEEyessubnumber\label{eq:Gbi1i1ij2}
\end{IEEEeqnarray}
Then, the effective channel matrix in \eqref{eq:u1K} can be written as $\left[\begin{array}{l}\bar{\mathbf{G}}_{11{,}j}{,}{\forall}j{=}2{,}\cdots{,}n \\ \bar{\mathbf{G}}_{l1{,}j}{,}{\forall}\{l{,}j\}{=}2{,}\cdots{,}n{,}l{\neq}j\end{array}\right]{=}\hat{\mathbf{H}}_1\hat{\mathbf{G}}_1{,}$ where
\begin{IEEEeqnarray}{rcl}
\hat{\mathbf{H}}_1&{=}&\text{\rm Bdiag}\left\{\left\{\left[\mathbf{H}_{1j}^{rt}\mathbf{C}_j\,\,\, {-}\mathbf{H}_{11}^{rt}\mathbf{C}_1\right]\right\}_{{\forall}j{=}2{,}\cdots{,}n}\quad
\left\{\left[\mathbf{H}_{lj}^{rt}\mathbf{C}_j\,\,\,{-}\mathbf{H}_{l1}^{rt}\mathbf{C}_1\right]\right\}_{ {\forall}\{l{,}j\}{=}2{,}\cdots{,}n{,}l{\neq}j} \right\}{,}\label{eq:Hhat}\\
\hat{\mathbf{G}}_1&{=}&{\rm stack}\left\{{\rm stack}\{\mathbf{D}_{1j}\mathbf{G}_{11}^{is}{,} \mathbf{D}_{j1}\mathbf{G}_{j1}^{is}\}_{{\forall}j{=}2{,}\cdots{,}n}{,}{\rm stack}\{\mathbf{D}_{lj}\mathbf{G}_{l1}^{is}{,} \mathbf{D}_{l1}\mathbf{G}_{l1}^{is}\}_{{\forall}\{l{,}j\}{=}2{,}\cdots{,}n{,}l{\neq}j}\right\}{.}\label{eq:Gbi1}
\end{IEEEeqnarray}

In \eqref{eq:Hhat}, the size of each submatrix is $Nt_2{\times}2\left[(n{-}1)N{-}(n{-}2)M\right]t_1$. Since $\mathbf{H}_{lj}^{rt}{,}{\forall}j{=}2{,}\cdots{,}n{,}{\forall}l{\in}1{,}\cdots{,}n{,}l{\neq}j$ is independent of $\mathbf{C}_j{,}{\forall}j{=}1{,}\cdots{,}n$, it follows that $\hat{\mathbf{H}}_1$ is full rank, $(n{-}1)^2\min\left\{Nt_2{,}2\left[(n{-}1)N{-}(n{-}2)M\right]t_1\right\}$, almost surely. In \eqref{eq:Gbi1}, there are $2(n{-}1)^2$ submatrices, each of which is of size $\left[(n{-}1)N{-}(n{-}2)M\right]t_1{\times}Mt_1$ and is full rank $\left[(n{-}1)N{-}(n{-}2)M\right]t_1$, almost surely. Let us look at the submatrices $\mathbf{D}_{1j}\mathbf{G}_{11}^{is}$ and $\mathbf{D}_{lj}\mathbf{G}_{l1}^{is}$, ${\forall}\{l{,}j\}{=}2{,}\cdots{,}n{,}l{\neq}j$. These $(n{-}1)^2$ submatrices are linear independent of each other because $\mathbf{D}_{lj}$ is related to the outgoing channels of Tx$j$ which is independent of $\mathbf{G}_{l1}^{is}$. However, the remaining $(n{-}1)^2$ blocks are equal to each other according to \eqref{eq:DG}, but they are independent of the other $(n{-}1)^2$ blocks. Consequently, the rank of $\hat{\mathbf{G}}_1$ is
\begin{IEEEeqnarray}{rcl}
\min\left\{\left((n{-}1)^2{+}1\right)\left[(n{-}1)N{-}(n{-}2)M\right]t_1{,}Mt_1\right\}{.}\label{eq:Gbrk}
\end{IEEEeqnarray}

To ensure the decodability, the rank of $\hat{\mathbf{G}}_1$ should be $Mt_1$. Hence, the inequality in \eqref{eq:constraints} is immediate according to \eqref{eq:Gbrk}. Given this condition, it can be verified that the rank of $\hat{\mathbf{H}}_1$ in \eqref{eq:Hhat} is $(n{-}1)^2Nt_2$. By setting $(n{-}1)^2Nt_2{=}Mt_1$, we can see that the effective channel matrix in \eqref{eq:u1K} is full rank almost surely because $\hat{\mathbf{G}}_1$ is independent of $\hat{\mathbf{H}}_1$.

\subsection{Proof of the linear independence of the observations in \eqref{eq:yKph2_effect}}
The derivation follows the footsteps of \cite[Appendix B]{Abdoli13}. Since performing a row transformation does not change the rank of a matrix, we replace the last $M$ rows of $\mathbf{G}_{k1}$ and $\mathbf{G}_{k2}$ by $\mathbf{F}_{k2}\mathbf{G}_{k1}$ and $\mathbf{F}_{k2}\mathbf{G}_{k2}$, respectively, where $\mathbf{F}_{k2}$ is such that $\mathbf{F}_{k2}\mathbf{G}_{k2}{=}\mathbf{0}$. Then, the effect channel matrix in \eqref{eq:yKph2_effect} rewrites as $\mathbf{Z}{=}\left[\begin{array}{cc}\tilde{\mathbf{G}}_{k1} & \tilde{\mathbf{G}}_{k2}\\ \mathbf{A}& \mathbf{0} \end{array}\right]$
where $\tilde{\mathbf{G}}_{k1}$ and $\tilde{\mathbf{G}}_{k2}$ are the first $Nt{-}M$ rows of $\mathbf{G}_{k1}$ and $\mathbf{G}_{k2}$, respectively. Clearly, $\tilde{\mathbf{G}}_{k1}$ and $\tilde{\mathbf{G}}_{k2}$ are full rank almost surely. Besides, $\mathbf{A}{\triangleq}{\rm stack}\left\{\{\mathbf{F}_{j2}\mathbf{G}_{j1}\}_{{\forall}j{=}k{,}m{+}1{,}m{+}2{,}\cdots{,}K}\right\}$ is of size $Mt{\times}Mt$.

As explained in \cite[Appendix B]{Abdoli13}, if $\mathbf{A}$ is full rank, then the matrix $\mathbf{Z}$ is full rank using \cite[Lemma 2]{Abdoli13}. Let us express $\mathbf{A}$ as $\mathbf{A}{=}\left[\mathbf{F}_2{\circ}\mathbf{H}_1\right]\mathbf{W}_1$, where $\circ$ denotes the block-wise product and $\mathbf{F}_2{=}{\rm stack}\left\{\mathbf{F}_{k2}{,}\mathbf{F}_{(m{+}1)2}{,}\cdots{,}\mathbf{F}_{K2}\right\}$ and $\mathbf{H}_1{=}{\rm stack}\left\{\mathbf{H}_{k1}{,}\mathbf{H}_{(m{+}1)2}{,}\cdots{,}\mathbf{H}_{K2}\right\}$.

Note that $\mathbf{F}_{j2}$ is the left null space of $\mathbf{G}_{j2}$, thus $\mathbf{F}_2$ is independent of $\mathbf{H}_1$ and $\mathbf{F}_2{\circ}\mathbf{H}_1$ is full rank $Mt_1$ almost surely. Moreover, as $\mathbf{W}_1$ is independent of $\mathbf{F}_2$ and $\mathbf{H}_1$, $\mathbf{A}$ is full rank almost surely, which completes the proof.

%% file: ICdelay_1col.bbl
\begin{thebibliography}{10}
\providecommand{\url}[1]{#1}
\csname url@samestyle\endcsname
\providecommand{\newblock}{\relax}
\providecommand{\bibinfo}[2]{#2}
\providecommand{\BIBentrySTDinterwordspacing}{\spaceskip=0pt\relax}
\providecommand{\BIBentryALTinterwordstretchfactor}{4}
\providecommand{\BIBentryALTinterwordspacing}{\spaceskip=\fontdimen2\font plus
\BIBentryALTinterwordstretchfactor\fontdimen3\font minus
  \fontdimen4\font\relax}
\providecommand{\BIBforeignlanguage}[2]{{%
\expandafter\ifx\csname l@#1\endcsname\relax
\typeout{** WARNING: IEEEtran.bst: No hyphenation pattern has been}%
\typeout{** loaded for the language `#1'. Using the pattern for}%
\typeout{** the default language instead.}%
\else
\language=\csname l@#1\endcsname
\fi
#2}}
\providecommand{\BIBdecl}{\relax}
\BIBdecl

\bibitem{Hao15MISOIC}
H.~Hao and B.~Clerckx, ``{D}egrees-of-freedom of the {K}-user {MISO}
  interference channel with delayed local {CSIT},'' in \emph{2015 IEEE
  International Conference on Communications (ICC)}, London, UK, June 2015.

\bibitem{JafarIA08}
V.~Cadambe and S.~Jafar, ``{I}nterference alignment and {D}egrees of {F}reedom
  of the {K}-user interference channel,'' \emph{IEEE Trans. on Infor. Theory},
  vol.~54, no.~8, pp. 3425--3441, Aug 2008.

\bibitem{GouKMN}
T.~Gou and S.~Jafar, ``{D}egrees of {F}reedom of the {K} user {M} x {N} {MIMO}
  interference channel,'' \emph{IEEE Trans. on Infor. Theory}, vol.~56, no.~12,
  pp. 6040--6057, Dec 2010.

\bibitem{Wang14SubspaceChain}
C.~Wang, T.~Gou, and S.~Jafar, ``{S}ubspace alignment chains and the {D}egrees
  of {F}reedom of the three-user {MIMO} interference channel,'' \emph{IEEE
  Trans. on Infor. Theory}, vol.~60, no.~5, pp. 2432--2479, May 2014.

\bibitem{JarfarMIMOIA2pair}
S.~Jafar and M.~Fakhereddin, ``Degrees of freedom for the mimo interference
  channel,'' in \emph{2006 IEEE Inter. Symp. on Infor. Theory}, July 2006, pp.
  1452--1456.

\bibitem{VV11a}
C.~Vaze and M.~Varanasi, ``{T}he {D}egree-of-{F}reedom regions of {MIMO}
  broadcast, interference, and cognitive radio channels with no {CSIT},''
  \emph{IEEE Trans. on Infor. Theory}, vol.~58, no.~8, pp. 5354 --5374, aug.
  2012.

\bibitem{HJSV09}
C.~Huang, S.~Jafar, S.~Shamai, and S.~Vishwanath, ``{O}n {D}o{F} region of
  {MIMO} networks without channel state information at transmitters,''
  \emph{IEEE Trans. on Infor. Theory}, vol.~58, pp. 849 --857, feb. 2012.

\bibitem{zhunoCSIT}
Y.~Zhu and D.~Guo, ``{T}he {D}egrees of {F}reedom of isotropic {MIMO}
  interference channels without state information at the transmitters,''
  \emph{IEEE Trans. on Inf. Theory}, vol.~58, no.~1, pp. 341--352, Jan 2012.

\bibitem{Ges12}
S.~Yang, M.~Kobayashi, D.~Gesbert, and X.~Yi, ``Degrees of freedom of time
  correlated {MISO} broadcast channel with delayed {CSIT},'' \emph{IEEE Trans.
  Inf. Theory}, vol.~59, no.~1, pp. 315--328, Jan. 2013.

\bibitem{Tse10}
M.~A. Maddah-Ali and D.~Tse, ``Completely stale transmitter channel state
  information is still very useful,'' \emph{IEEE Trans. Inf. Theory}, vol.~58,
  no.~7, pp. 4418--4431, July 2012.

\bibitem{Bruno15}
B.~Clerckx and D.~Gesbert, ``{S}pace-{T}ime encoded {MISO} broadcast channel
  with outdated {CSIT}: An error rate and diversity performance analysis,''
  \emph{IEEE Trans. on Commun.}, vol.~63, May 2015.

\bibitem{Mingbo15}
M.~Dai and B.~Clerckx, ``{T}ransmit beamforming for {MISO} broadcast channels
  with statistical and delayed {CSIT},'' \emph{IEEE Trans. on Commun.},
  vol.~63, April 2015.

\bibitem{TorrellasK}
M.~Torrellas, A.~Agustin, and J.~Vidal, ``On the degrees of freedom of the
  k-user {MISO} interference channel with imperfect delayed {CSIT},'' in
  \emph{Proc. of 2014 IEEE International Conference on Acoustics, Speech and
  Signal Processing (ICASSP)}, Florence, May 2014, pp. 1155--1159.

\bibitem{VVICdelay}
C.~S. Vaze and M.~K. Varanasi, ``The {DoF} region and interference alignment
  for the {MIMO} interference channel with delayed {CSIT},'' \emph{IEEE Trans.
  on Inf. Theory}, vol.~58, no.~7, pp. 4396--4417, July 2012.

\bibitem{xinping_miso_ic}
X.~Yi, D.~Gesbert, S.~Yang, and M.~Kobayashi, ``On the dof of the
  multiple-antenna time correlated interference channel with delayed csit,'' in
  \emph{2012 Conference Record of the Forty Sixth Asilomar Conference on
  Signals, Systems and Computers (ASILOMAR)}, Nov 2012, pp. 1566--1570.

\bibitem{Ghasami11}
A.~Ghasemi, A.~S. Motahari, and A.~K. Khandani, ``Interference alignment for
  the mimo interference channel with delayed local csit,'' \emph{available on
  arXiv: http://arxiv.org/pdf/1102.5673v1.pdf}, 2011.

\bibitem{Abdoli13}
M.~J. Abdoli, A.~Ghasemi, and A.~K. Khandani, ``On the degrees of freedom of
  k-user {SISO} interference and {X} channels with delayed {CSIT},'' \emph{IEEE
  Trans. on Inf. Theory}, vol.~59, no.~10, pp. 6542--6561, Oct 2013.

\bibitem{MalekiRIA}
H.~Maleki, S.~Jafar, and S.~Shamai, ``Retrospective interference alignment over
  interference networks,'' \emph{IEEE Journal of Selected Topics in Signal
  Processing}, vol.~6, no.~3, pp. 228--240, June 2012.

\bibitem{Maggi12}
L.~Maggi and L.~Cottatellucci, ``Retrospective interference alignment for
  interference channels with delayed feedback,'' in \emph{Wireless
  Communications and Networking Conference (WCNC), 2012 IEEE}, April 2012, pp.
  453--458.

\bibitem{TorrellasMNK}
M.~Torrellas, A.~Agustin, and J.~Vidal, ``Dof-delay trade-off for the k-user
  mimo interference channel with delayed {CSIT},'' on arXiv: 1504.05498, 2015.

\bibitem{xinping_mimo}
X.~Yi, S.~Yang, D.~Gesbert, and M.~Kobayashi, ``The degrees of freedom region
  of temporally correlated mimo networks with delayed csit,'' \emph{IEEE Trans.
  on Infor. Theory}, vol.~60, no.~1, pp. 494--514, Jan 2014.

\end{thebibliography}
